\documentclass[journal]{IEEEtran}
\usepackage{cite}
\usepackage{longtable}

\ifCLASSINFOpdf
   \usepackage[pdftex]{graphicx}
   \graphicspath{{../pdf/}{../jpeg/}}
\else
\fi
\usepackage[cmex10]{amsmath}
\usepackage{algorithmic}

\usepackage{array}

\usepackage{stfloats}
\usepackage{url}

\usepackage{soul,xcolor}
\usepackage{url}

\begin{document}
\setstcolor{red}
%
\title{A Comprehensive Survey of Control Strategies for Autonomous Quadrotors}
%
%
%

\author{Jinho~Kim,~\IEEEmembership{Student Member,~IEEE,}
        S.~Andrew~Gadsden,~\IEEEmembership{Member,~IEEE,}
        and~Stephen~A.~Wilkerson,~\IEEEmembership{Member,~IEEE,}
\thanks{Jinho Kim is with the Department
of Mechanical Engineering, University of Maryland, Baltimore County, Baltimore,
MD, 21250 USA (e-mail: umbcjhkim@umbc.edu).}
\thanks{S. Andrew Gadsden is with the Department of Mechanical Engineering, University of Guelph, Guelph, Ontario, N1G 2W1 Canada (e-mail: gadsden@uoguelph.ca)}
\thanks{Stephen A. Wilkerson is with the Department of Engineering and Computer Science, York College of Pennsylvania, York, PA, 17403 USA (e-mail: swilkerson@ycp.edu)}
}

\maketitle

\begin{abstract}
Over the past several decades there has been a constant increase in the use of Unmanned Aerial Systems (UAS). Hence, there has also been a growth in the number of control algorithms to service the many applications embodied by these vehicles. Initially UAS were made popular by the military for Reconnaissance, Intelligence, Surveillance, and Target Acquisition (RISTA) applications. Now-a-days UAS are used for everything from crop surveys to tourism. Nowhere is this more evident than with multi-rotor Unmanned Aerial Vehicle (UAV). This paper presents a survey of control methods for multi-rotor systems, namely quadrotors. In doing so, we hope to visualize a clear path to what additional capabilities might be needed in the future. In our examination, we review many of the notable research organizations and their efforts to expand the utility of multi-rotor aircraft. We also summarize the basic literature definitions and control strategies for autonomous quadrotors.
\end{abstract}


%
\IEEEpeerreviewmaketitle

\section{Introduction} \label{intro}
%
%
%
%
\IEEEPARstart{U}{nmanned} Aerial Systems (UAS) have grown significantly in popularity and notoriety in the past few years. Moreover, there has been an explosion of new UAS applications for more than a decade now. The initial purpose for UAS was military Reconnaissance, Intelligence, Surveillance, and Target Acquisition (RISTA) applications \cite{glade2000unmanned, lyon2004military, samad2007network, chen2010uav}. Furthermore, numerous civilian applications have now been developed to include: first responder disaster response, education, environment and climate studies, tourism, mapping, crop assessments, weather, traffic monitoring and many others \cite{quigley2005target, rodriguez2006emergency, ahmadzadeh2006multi, heintz2007images, hirokawa2007small, nebiker2008light, jensen2009aggieair, techy2009uav, debusk2010unmanned, goktougan2009rotary, casbeer2006cooperative, maza2011experimental, samad2013potential, ezequiel2014uav}.

\begin{figure}
\centering
\includegraphics[width=0.45\textwidth]{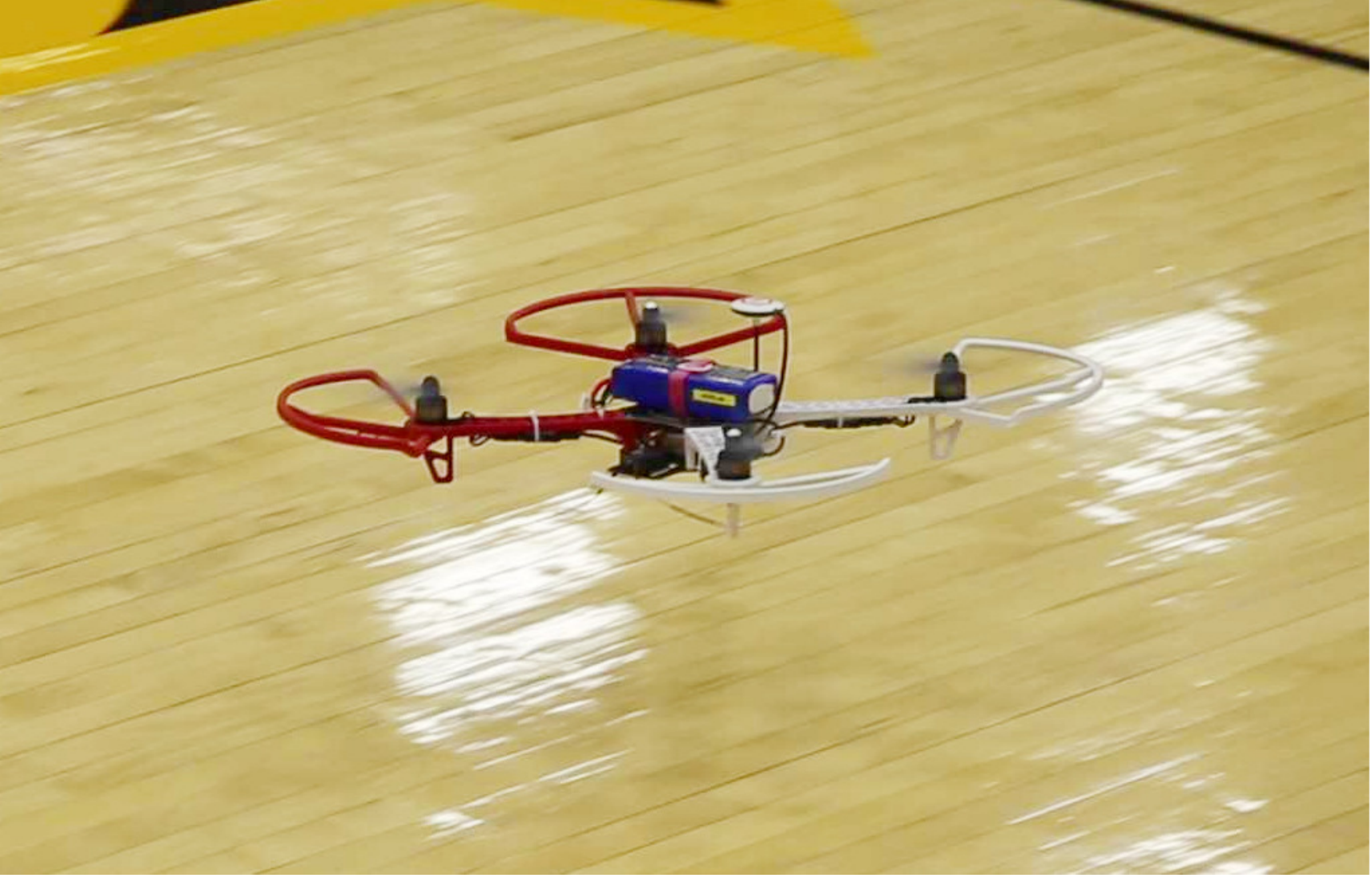}
\caption{A quadrotor made by DJI (F450) \cite{dji} with propeller guards.}
\label{fig_quadrotor}
\end{figure}

Among the types of UAS, a quadrotor, consists four rotors fixed to a rigid cross frame as shown in Fig.~\ref{fig_quadrotor}, is one of the most popular UAS. Because the quadrotor has a number of advantages over a fixed-wing aircraft due to its ability to take-off and land vertically. Furthermore, like a helicopter, the quadrotor can hover, but with its four rotors it is capable of lifting larger payloads relative to it's own weight. Furthermore, a small-sized quadrotor is agile, highly maneuverable, and is inherently more stable due to the four rotor design with counter rotating props eliminating the need for a tail rotor. Owing to these advantages, and a growing-range of useful applications, the quadrotor has become a popular subject for research. In addition to scientific studies, quadrotors are being developed for commercial ventures which are expected to expand rapidly during the next decade \cite{dillow2014get}.

\begin{figure*}[t]
\centering
\includegraphics[width=15cm, height=10cm]{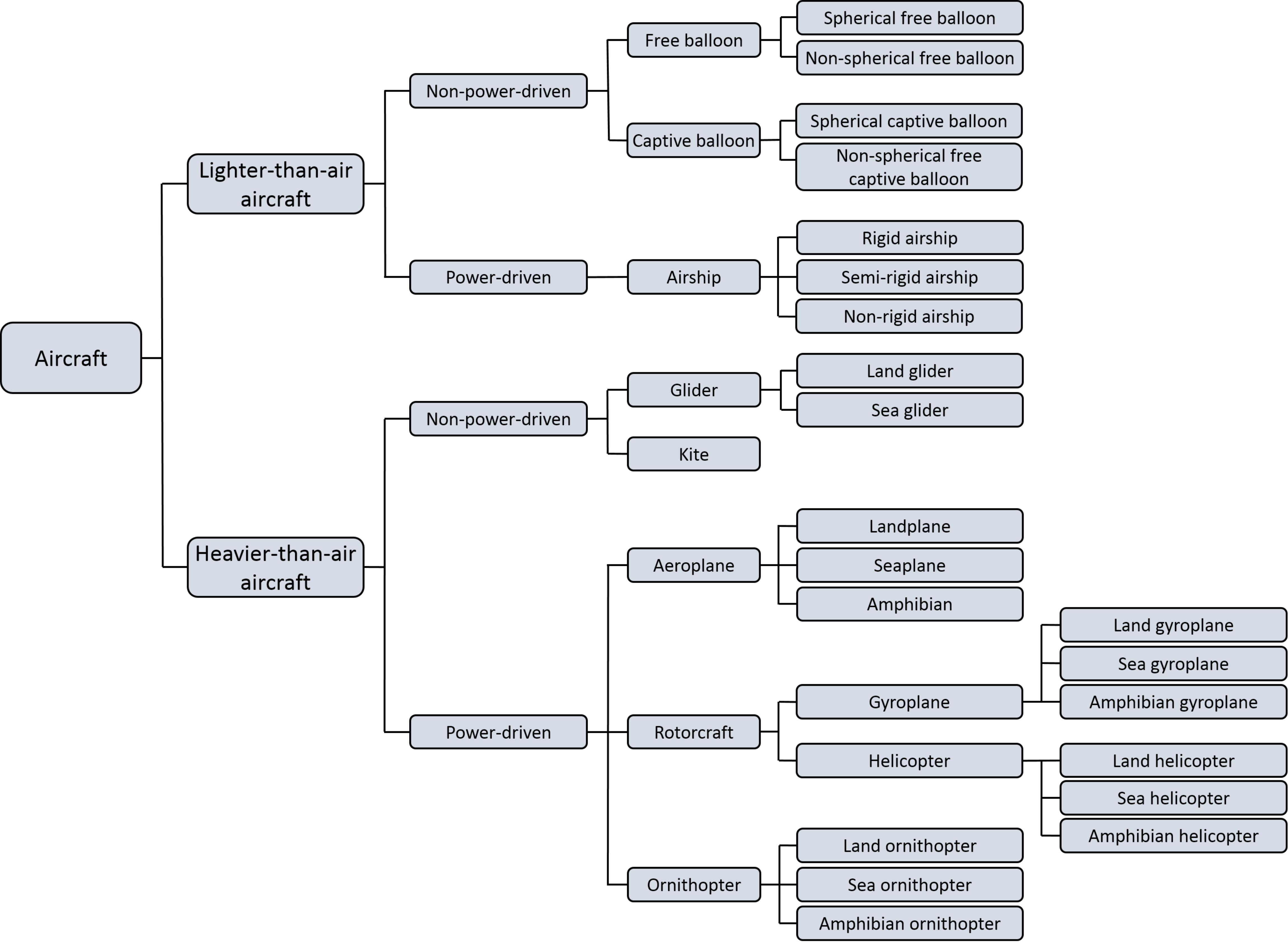}
\caption{ICAO Classification of aircraft.}
\label{fig_classification}
\end{figure*}

Numerous research groups have developed aerial vehicles to include fixed-wing aircraft, helicopters, and multi-rotor designs enabling them to operate autonomously. Studies of autonomous flight have led to advances in control theory, and have contributed significantly to the available body of literature. Furthermore, advances in brushless motors and lightweight electronics have led to smaller and more agile quadrotors which is shown in Fig.~\ref{fig_quadrotor}. Since the early turn of the century quadrotors have become a key figure or staple of UAV research and development. Much of the rise in quadrotor popularity has been a result of the adaptability of the platform for specific tasks that were formerly deemed difficult or impossible to do. Much of the early work on quadrotor developments focused on the reliable control of the system. However, more recently studies offer a wide range of topics and uses. In particular, this survey paper will focus on the development of control strategies used to improve quadrotors capabilities. Due to the growing body of quadrotor research, this survey paper seems prudent to overview essential research results on control strategies for the quadrotor.

This paper is organized as follows. \ref{definitions} defines primary terminologies and introduces some of the notable research groups that study and develop autonomous quadrotors. \ref{control} provides a broad review of quadrotor control methodologies. This section is broken into three subsections to include: linear and robust controllers, nonlinear controllers, and intelligent controllers.

\section{Definitions and Research Groups} \label{definitions}
Prior to review of quadrotor control methods, it is pragmatic to establish some basic definitions, classifications and a body of research groups from which to sample. Hence, this section gives some general classifications and definitions of aircrafts. Afterwards, we introduce some of the research groups focused on quadrotor research.

\subsection{Classification of aircrafts and primary definitions} \label{classification}
In this section, we introduce general classification of aircrafts suggested by the International Civil Aviation Organization (ICAO) and define some terminologies. Fig. \ref{fig_classification} shows a classification of aircraft proposed by ICAO, and it is classified depending on not only the weight of the aircraft, but also flying principle and propulsion method \cite{icao2003annex}. From this figure, we can see the quadrotor belongs to rotorcraft. Now, we propose some definitions for this paper. The proposed definitions are based on US Public Law 112-95 \cite{faa2012public}, and modified to be suitable for the quadrotor.

\textbf{Definition 1.} This paper uses the word quadrotor, not quadcopter. Since the prefix $quad$ and $rotor$ are from a Latin root whereas $copter$ is derived from a Greek root. Therefore, the word quadrotor seems more appropriate than quadcopter from a linguistic point of view \cite{mahony2012multirotor}. Hence, we use the term quadrotor throughout this paper. This is intended to designate a craft that has four independent rotors at the extremities of the cross-shaped body similar to what can be seen in Fig. \ref{fig_quadrotor}.
 
\textbf{Definition 2.} An Unmanned Aerial Vehicle (UAV) is a reference to just the aircraft. However, Unmanned Aircraft Systems (UAS) include the ground control and communications systems too. Generally speaking, the term for UAS includes vehicles that are controlled either manually by a ground pilot or autonomously by onboard sensors and computers. Nonetheless both have onboard controllers to facilitate the control. Such vehicles can be used for missions such as following predesigned paths or used to explore unknown environments with some Guidance, Navigation and Control (GNC) algorithms. Hence, we use the term UAS throughout to define a Remotely Piloted Aircraft (RPA) that has a control system associated with it.
 
\textbf{Definition 3.} With definition 2, we define a RUAS for Rotorcraft UAS and use the term ARUAS for Autonomous RUAS. Specially, we use the term Autonomous Quadrotor (AQ) for the quadrotor which flies autonomously without any manual control in this paper. 

\subsection{Quadrotor research groups} \label{groups}
In this section, we present some of the groups which have researched autonomous flight of quadrotors. The quadrotor is very well commercialized now and these vehicles and control systems are being researched and used all over the world. Among the numerous groups doing research in this area, we summarize some of the efforts relative to our interests in Table \ref{table_research group}.

\begin{table*}
    \caption{Major Research Groups}
    \label{table_research group}
    \begin{tabular}{ | p{3.5cm} | p{3.5cm} | p{6.5cm} | p{3.0cm} | }
    
    \hline
    Name of Group & Research Areas & Achievement & Quadrotor Platforms \\ \hline

General Robotics, Automation, Sensing and Perception (GRASP) Lab., University of Pennsylvania, US. \newline
 \url{(https://www.grasp.upenn.edu/research-groups/kumar-lab)} &
- Flying gripper \newline
- Autonomous Robotic Rotorcraft for Exploration, Surveillance and Transportation (ARREST) \newline
- Aerial robots for remote autonomous exploration and mapping \newline
- Printable robots \newline
- Autonomous micro UAVs \newline
- Vision-based control &
- Navigate in complex, 3D environments with or without GPS to include Exploration of the fully autonomous quadrotor in disaster areas and 3D mapping \newline
- Creating printable, small, autonomous quadrotors \newline
- Motion planning algorithms for large teams of quadrotors with dynamic constraints \newline
- Development of trajectories and controllers enabling a quadrotor aggressive maneuvers \newline
- Cooperation using multiple quadrotors \newline
- Develop a nonlinear vision-based controller for trajectory tracking &
- AscTec Hummingbird \newline
- AscTec Pelican \newline
- KMel Nano \newline
- KMel kQuad 500 \newline
- Crazyflie quadrotor \newline
- Customized qudrotor \\  \hline


UAV Research Facility, Georgia Institute of Technology, US.
\url{(http://www.uavrf.gatech.edu/)} &
- Avionics integration  \newline
- GNC algorithm  \newline
- SLAM algorithm using a visual sensor &
- Exploration of indoor areas using fully autonomous quadrotors with on-board chipsets  \newline
- Position and heading estimation using a scanning laser rangefinder and SLAM algorithm  \newline
-  Monocular EKF-SLAM navigation system &
- AscTec Pelican \\ \hline

Robust Robotics Group, MIT, US.
\url{(http://groups.csail.mit.edu/rrg/index.php?n=Main.Micro-AirVehicleNavigationAndControl)} &
- Exploration in GPS-denied Environments \newline
- SLAM algorithms &
- Visual odometry and laser scan matching GPS-denied flight with onboard the vehicle \newline
- Autonomous flight using SLAM algorithms in unknown indoor environments &
- AscTec Pelican \\ \hline

CyPhy Lab., Queensland University of Technology, Australia.
\url{(https://wiki.qut.edu.au/display/cyphy/Robotics@QUT)} &
- Image based Visual control with autonomous flight in GPS-impaired environment  \newline
- Open source palm sized quadrotor	&
- Monocular vision based autonomous navigation  \newline
- Visual SLAM using visual expectation  \newline
- Collision avoidance method using spherical image based visual servoing  \newline
- Position based visual servoing for pole inspection task  \newline
- Palm size open source micro quadrotor constructed on a single Printed Circuit Board	&
- MikroKopter  \newline
- Parrot AR.Drone \\ \hline

Institute for Dynamic Systems and Control, ETHZ, Switzerland.
\url{(http://www.idsc.ethz.ch/research-dandrea.html)} &
- High-precision maneuvers and flight control  \newline
- Physical interaction and novel sensing modalities \newline
- Cooperation of quadrotors	&
- Optimal and learning controller for aggressive maneuvers  \newline
- Algorithm for generating flight trajectories  \newline
- Cooperation of multiple quadrotors  \newline
- Fault-tolerant control &
- AscTec Hummingbird  \newline
- Customized quadrotor \\ \hline

Autonomous Systems Lab. (ASL), ETHZ, Switzerland.
\url{(http://www.asl.ethz.ch/)} &
- Flight control \newline
- Collaborative Aerial Robotic Workers \newline
- Search and Rescue Robotic Platform \newline
- Flourish - Aerial Data Collection and Analysis for Precision Farming	&
- Nonlinear controller for attitude, altitude and position control to include Obstacle avoidance 
controller using four ultrasound sensors  \newline
- Tele-operated quadrotor with altitude and position control  and autonomous control
 for attitude control against model uncertainties and external disturbances &
- Customized quadrotor \\ \hline

Computer Vision Group, Technical University of Munich, Germany.
\url{(https://vision.in.tum.de/)} &
- Vision-based navigation and mapping  \newline
- Nano-copters	&
- Nano-quadrotor with wireless video capability  \newline
- Visual-inertial SLAM using onboard camera  \newline
- Acquiring textured 3D model of indoor spaces using the RGB-D camera  \newline
- Person following and gesture recognition using an on-board depth camera	&
- Parrot AR.Drone  \newline
- Crazyflie quadrotor \\ \hline

Intelligent Control System Lab. (ICSL), SNU, South Korea.
\url{(http://icsl.snu.ac.kr/)} &
- Flight control  \newline
- Vision-based control and estimation  \newline
- Aerial manipulation  \newline
- Cooperation of quadrotors	&
- Nonlinear controllers for an autonomous quadrotor  \newline
- Robust controllers using TS (Takagi-Sugeno) fuzzy model and an LMI (Linear Matrix Inequality) approach  \newline
- Integrate adaptive image-based visual servoing with adaptive sliding mode control  \newline
- Vision-based lifting of a payload using two quadrotors  \newline
- Onboard control of a small quadrotor using an optical flow sensor  \newline
- Aerial manipulation using a quadrotor with a two-DOF robot arm	&
- Xcopter Smart Xcopter  \newline
- Mikrokopter MK-hexa2  \newline
- Ascending Technologies Firefly hexacopter  \newline
- Customized quadrotor 	\\ \hline

Unmanned System Research Group, USRG, KAIST, South Korea.
\url{(http://unmanned.kaist.ac.kr/)} &
- Vision-based control  \newline
- Indoor 3D navigation   &
- Indoor flight control using outboard multi-camera visual feedback  \newline
- Vision-based target detection and following algorithm using color and image moment  \newline
- Outdoor autonomous landing on a moving platform using an omnidirectional camera	&
- Parrot AR.Drone  \newline
- Aeryon Scout \newline
- DJI F450  \newline
- Customized quadrotor	\\ \hline

    \end{tabular}
\end{table*}

\begin{figure*}[t]
\centering
\includegraphics[width=15cm, height=10cm]{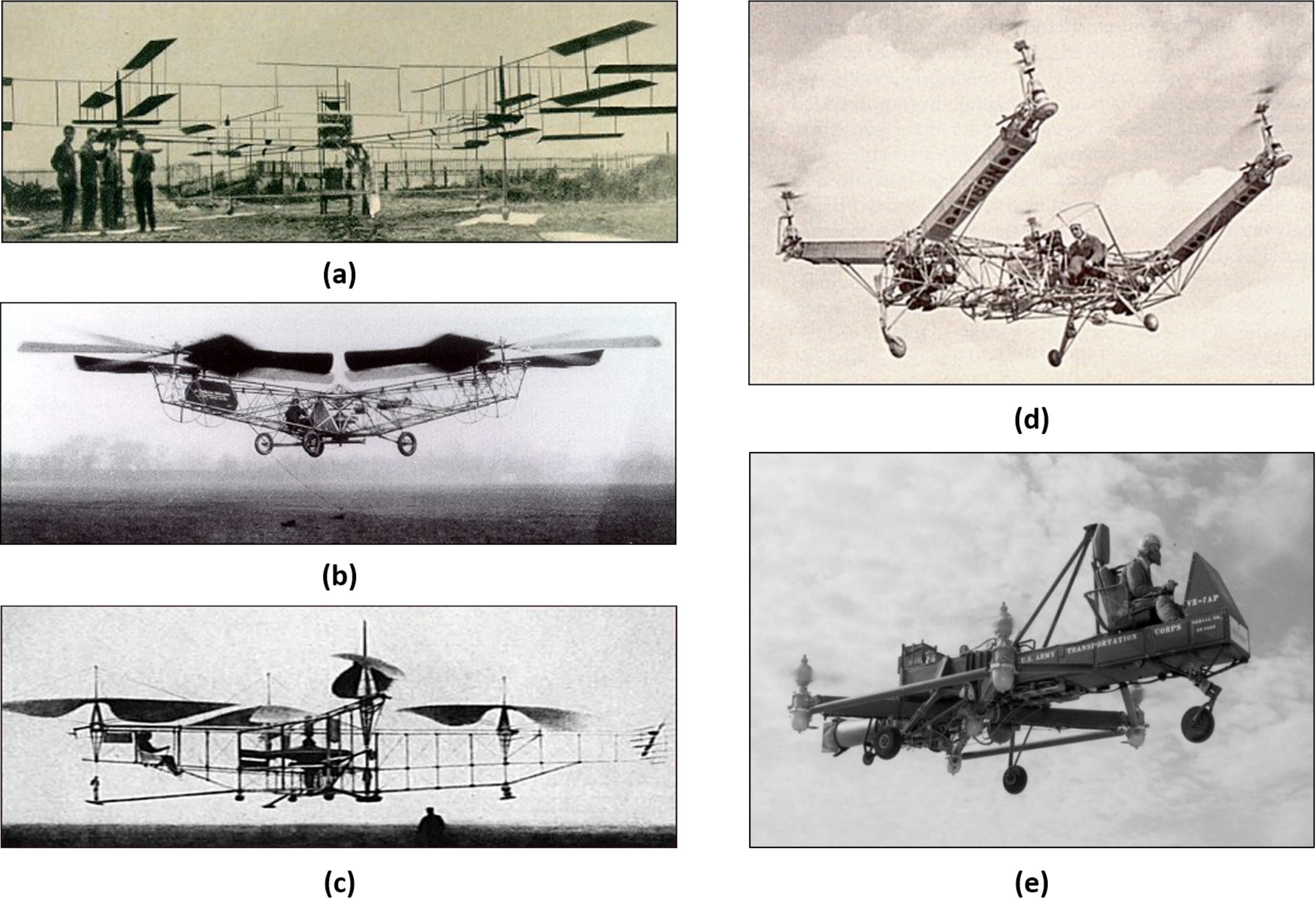}
\caption{History of quadrotor: (a) Br\`{e}guet-Richet Gyroplane No. 1; (b) Oehmichen No.2; (c) Bothezat helicopter; (d) Convertawings Model A; (e) Curtiss-Wright VZ-7}
\label{fig_quad_history}
\end{figure*}

\subsection{History of Quadrotors} \label{history}
Since Orville and Wilbur Wright achieved the first controlled-powered human flight of a heavier-than-air machine in 1903, a huge number of aerial vehicles have been invented, tested, and flown. Most of the early developments were achieved using the Wright fixed-wing design \cite{john1989introduction}. However, one of the major drawbacks of the fixed-wing design is the requirement for runways needed for landing and take-offs. In order to overcome this restriction, rotary-wing aerial vehicles were introduced in various configurations.

\subsubsection{Early attempts} \label{early}
The first quadrotor-type design was introduced in 1907 by the Br\`{e}guet brothers, and was referred to as the Br\`{e}guet-Richet Gyroplane No. 1, Fig. \ref{fig_quad_history}(a) \cite{young1982helicopters}. The design of this vehicle was very unique. Each rotor had four biplane-type blades (two deep) which provided a total of 32 separate lifting surfaces, and was connected with an Internal Combustion Engine (ICE) through a belt and pulley transmission system. The four rotors were configured into two clockwise and counter-clockwise rotating pairs; thereby canceling torsion effects on the body frame. This standard design principle is used in modern quadrotors today. Note, however, that the Br\`{e}guet-Richet Gyroplane No. 1 did not have any control surfaces for maneuvering rendering it impractical. 

\'{E}tienne Oehmichen further experimented and designed a variety of quadrotors. The Oehmichen No. 2 had two-bladed rotors that were located at the end of the frame as shown in Fig. \ref{fig_quad_history}(b). These blades were able to be warped, thereby modifying the blade angle of attack giving the vehicle limited control. Two propellers were located at the nose of the vehicle, and were used for yaw control. The Oehmichen No. 2 is considered to be a mixed-design between a quadrotor and helicopter. In 1922, George de Bothezat and Ivan Jerome developed the quadrotor structure further by including six-bladed rotors and two additional propellers, Fig. \ref{fig_quad_history}(c). Two small propellers were mounted for thrust and yaw control. Collective pitch control was also applied on this vehicle \cite{de1926helicopter}. Although there was initial interest and research in quadrotors, the design became less popular during the following two decades due to weight and technical problems. 

In 1956, the Convertawings Model A quadrotor was designed and introduced as shown in Fig. \ref{fig_quad_history}(d). This unique design utilized two engines that operated four rotors for lift. Flight tests successfully demonstrated and proved the ability of forward flight with this design. However, due to a lack of orders, this model was shortly abandoned. In 1958, the Curtiss-Wright company designed the Curtiss-Wright VZ-7 for the US Army, Fig. \ref{fig_quad_history}(e). This quadrotor model had four rotors and implemented individual speed controllers for each rotor. The Curtiss-Wright VZ-7 design may be considered a precursor to modern quadrotor designs. Quadrotor research has accelerated in the new Millennium due to the advances in computing, sensors, power and energy and manufacturing. New uses for these vehicles are turning up at a growing pace.

\subsubsection{Recent trend} \label{recent}
Quadrotor designs have now been utilized to develop small-scale UAV. Numerous universities and companies have developed research centers in an effort to improve the designs and applications of quadrotor. The UAV industry is expected to grow to \$22.4 billion by 2020 \cite{dillow2014get}. This growth has led rise to a number of companies offering commercial quadrotors at the 2015 International Consumer Electronics Show (CES) held in Las Vegas, Nevada \cite{metz2015ces}.

\section{Quadrotor Dynamics}
\begin{figure}[t]
\begin{center}
\includegraphics[width=0.45\textwidth]{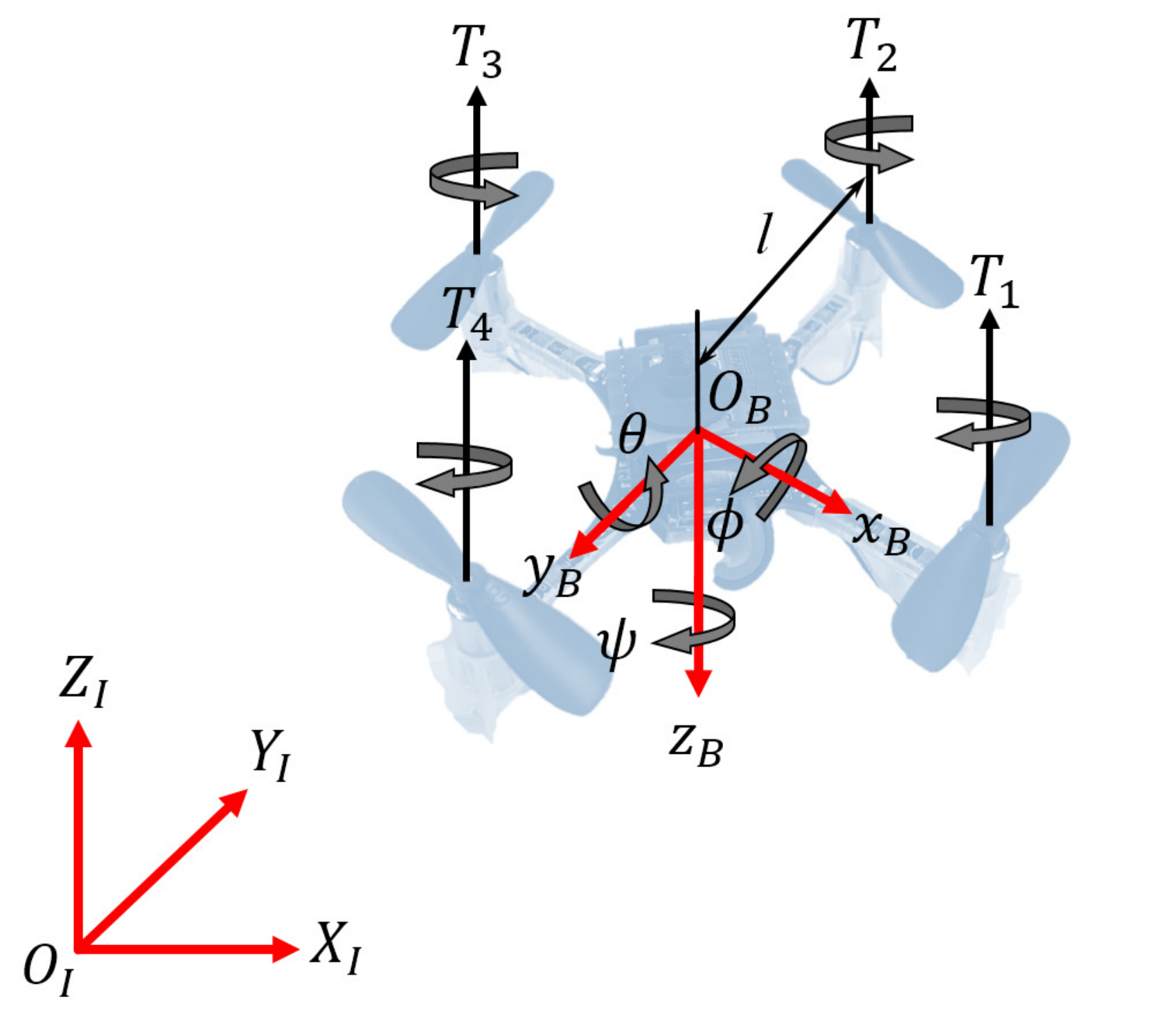}
\end{center}
\caption{A configuration of a quadrotor. [$X_I$, $Y_I$, $Z_i$] denotes the inertial coordinate frame, and [$x_B$, $y_B$, $z_B$] is the body coordinate frame. $[x, y, z]$ denotes the position of the quadrotor in the inertial frame, and $[\phi, \theta, \psi]$ represent roll, pitch, and yaw angles, respectively defined in the body frame. $T_i \; (i=1,2,3,4)$ is the thrust force of the $i$-th rotor, $l$ denotes the length between each rotor and the geometric center of the quadrotor.}
\label{fig:quad_quad}
\end{figure}

The quadrotor considered in this paper consists of a rigid cross frame and four fixed rotors which has fixed-pitch-angle blades, unlike classic helicopters that have variable-pitch-angle blades. The quadrotor is an under-actuated system with four inputs and six outputs, and the states are highly coupled. To control this system, modeling approaches are presented in \cite{altug2002control,bouabdallah2004design,goodman2015system}. The quadrotor can be controlled by varying thrust forces and moments generated by each rotor. To move along the $x$-axis, the quadrotor has to create a pitch angle ($\theta$) by varying the speed of rotors $1$ and $3$ while attempting to maintain the speeds of rotor $2$ and $4$ in Fig.~\ref{fig:quad_quad}. This is due to the fact that the $x$-translational motion is related to the tilt of the body frame with respect to the $y$-axis. Similarly, the quadrotor can fly along the $y$-axis due to the rotation of the frame $x$-axis, which generates a roll angle ($\phi$). In order to generate a yaw angle ($\psi$) with respect to the $z$-axis, the speeds of rotors $1$ and $3$ have to be increased while the speeds of rotor $2$ and $4$ are proportionally decreased.

To make the dynamics of quadrotor simple, it is assumed that the quadrotor is symmetric with respect to the $x$ and $y$ axes. Then, the center of gravity aligns with the geometric center of the quadrotor. The length between each rotor and the geometric center of the quadrotor is defined by $l$. The thrust forces, perpendicular to the $x$-$y$ plane, are generated by the four rotors and are defined by $T_i \; (i=1,2,3,4)$.

The dynamic equations of the quadrotor model without consideration of air drag can be presented as below \cite{altug2002control}.

\begin{equation} \label{eq:cssn_quad_dyn01}
\ddot{x} = \dfrac{1}{m}(T_{1}+T_{2}+T_{3}+T_{4})(\cos{\phi}\sin{\theta}\cos{\psi}+\sin{\phi}\sin{\psi}),
\end{equation}

\begin{equation} \label{eq:cssn_quad_dyn02}
\ddot{y} = \dfrac{1}{m}(T_{1}+T_{2}+T_{3}+T_{4})(\cos{\phi}\sin{\theta}\sin{\psi}-\sin{\phi}\cos{\psi}),
\end{equation}

\begin{equation} \label{eq:cssn_quad_dyn03}
\ddot{z} = \dfrac{1}{m}(T_{1}+T_{2}+T_{3}+T_{4})(\cos{\phi}\cos{\theta})-g,
\end{equation}

\begin{equation} \label{eq:cssn_quad_dyn04}
\ddot{\phi} = \dfrac{l}{J_{x}}(T_{2}-T_{4}),
\end{equation}

\begin{equation} \label{eq:cssn_quad_dyn05}
\ddot{\theta} = \dfrac{l}{J_{y}}(T_{1}-T_{3}),
\end{equation}

\begin{equation} \label{eq:cssn_quad_dyn06}
\ddot{\psi} = \dfrac{C}{J_{z}}(-T_{1}+T_{2}+T_{3}-T_{4}).
\end{equation}\\
\noindent Here, $[x, y, z]$ denotes the position of the quadrotor in the inertial frame, and $[\phi, \theta, \psi]$ represent roll, pitch, and yaw angles, respectively defined in the body frame. $m$ is the mass of the quadrotor, $J_{x,y,z}$ is the moments of inertia with respect to the axes. $C$ denotes the force-to-moment scaling factor and $g$ is the acceleration of gravity.

In order to simplify Eqs.~\ref{eq:cssn_quad_dyn01}-\ref{eq:cssn_quad_dyn06}, the control input terms $u_{i}$ are defined as Eqs.~\ref{eq:cssn_quad_u1}-\ref{eq:cssn_quad_u4}. $u_{1}$ is the control input of the total thrust, and $u_{2}$, $u_{3}$, and $u_{4}$ correspond to the control inputs of roll, pitch, and yaw moments, respectively.
\begin{equation} \label{eq:cssn_quad_u1}
u_1 = \dfrac{1}{m}(T_{1}+T_{2}+T_{3}+T_{4}),
\end{equation}

\begin{equation} \label{eq:cssn_quad_u2}
u_2 = \dfrac{1}{J_{x}}(T_{2}-T_{4}),
\end{equation}

\begin{equation} \label{eq:cssn_quad_u3}
u_3 = \dfrac{1}{J_{y}}(T_{1}-T_{3}),
\end{equation}

\begin{equation} \label{eq:cssn_quad_u4}
u_4 = \dfrac{C}{J_{z}}(-T_{1}+T_{2}+T_{3}-T_{4}).
\end{equation}\\
\noindent Then the dynamic equations can be represented as Eqs.~\ref{eq:cssn_quad_final_dyn01}-\ref{eq:cssn_quad_final_dyn06}.
\begin{equation} \label{eq:cssn_quad_final_dyn01}
\ddot{x} = u_{1}(\cos\phi\sin\theta\cos\psi+\sin\phi\sin\psi),
\end{equation}

\begin{equation} \label{eq:cssn_quad_final_dyn02}
\ddot{y} = u_{1}(\cos\phi\sin\theta\sin\psi-\sin\phi\cos\psi),
\end{equation}

\begin{equation} \label{eq:cssn_quad_final_dyn03}
\ddot{z} = u_{1}(\cos\phi\cos\theta)-g,
\end{equation}

\begin{equation} \label{eq:cssn_quad_final_dyn04}
\ddot{\phi} = u_{2}l,
\end{equation}

\begin{equation} \label{eq:cssn_quad_final_dyn05}
\ddot{\theta} = u_{3}l,
\end{equation}

\begin{equation} \label{eq:cssn_quad_final_dyn06}
\ddot{\psi} = u_{4}.
\end{equation}

\section{Control Strategies for a Quadrotor} \label{control}

\begin{figure}[!t]
\centering
\includegraphics[width=0.48\textwidth]{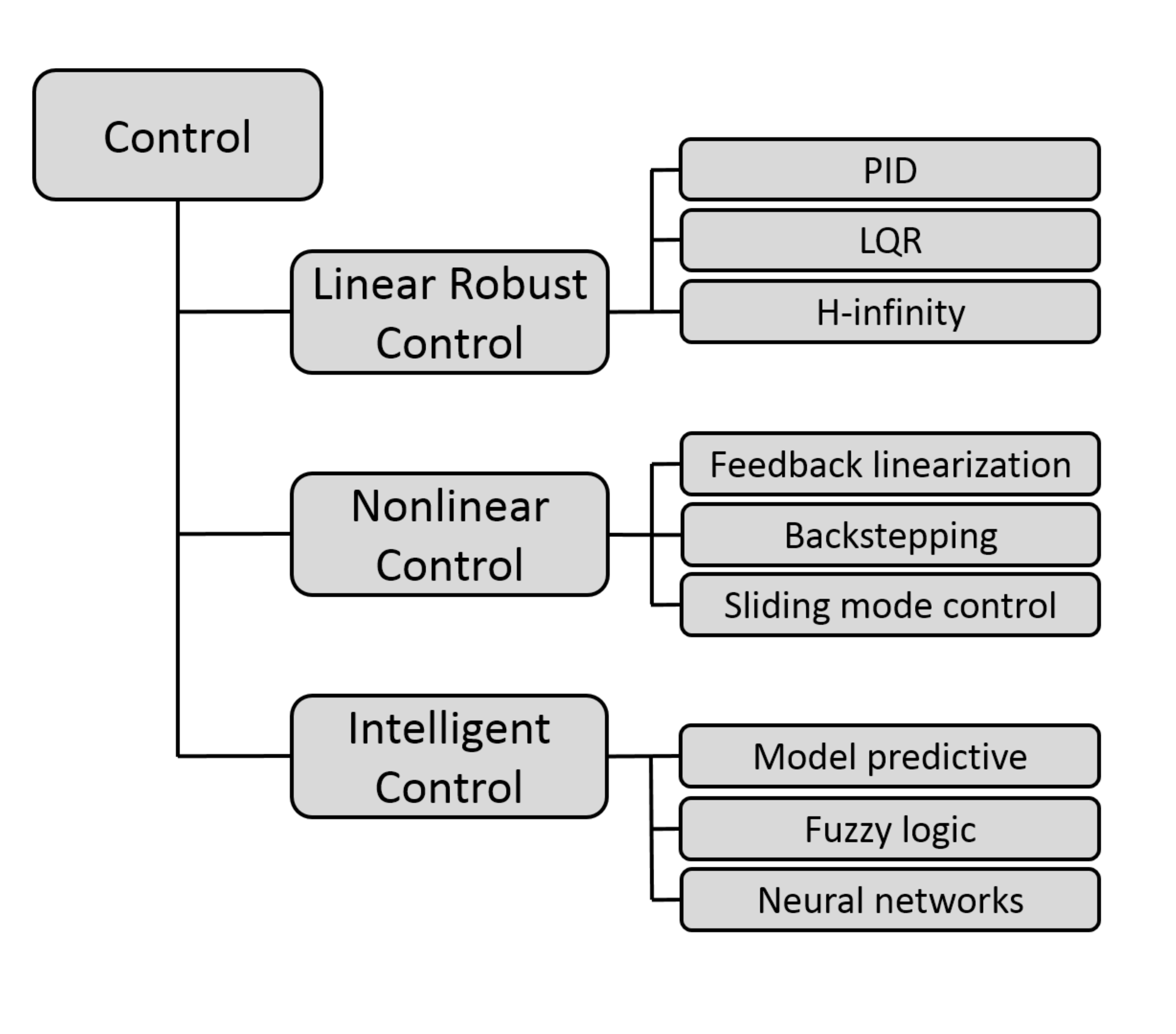}
\caption{Categorization of controllers.}
\label{fig_controller}
\end{figure}

In the past decades several different classes of controllers have been employed for the control of quadrotor. We break these into three categories as shown in Fig. \ref{fig_controller}. Then, within each category, we examine several of the more popular controllers and the contributions made by various researcher organizations.

\subsection{Linear Robust Controllers} \label{linear}
Early in quadrotor development it was found that linear controllers were sufficient to obtain stable flight. We examine several of these control techniques to include, a Proportional Integral Derivative (PID) controller, Linear Quadratic Controllers, and $H_{\infty}$ controller. Additionally, in this section we summarize some of the available literature dealing with linear controllers used specifically for quadrotors.

\begin{figure}[!t]
\centering
\includegraphics[width=0.48\textwidth]{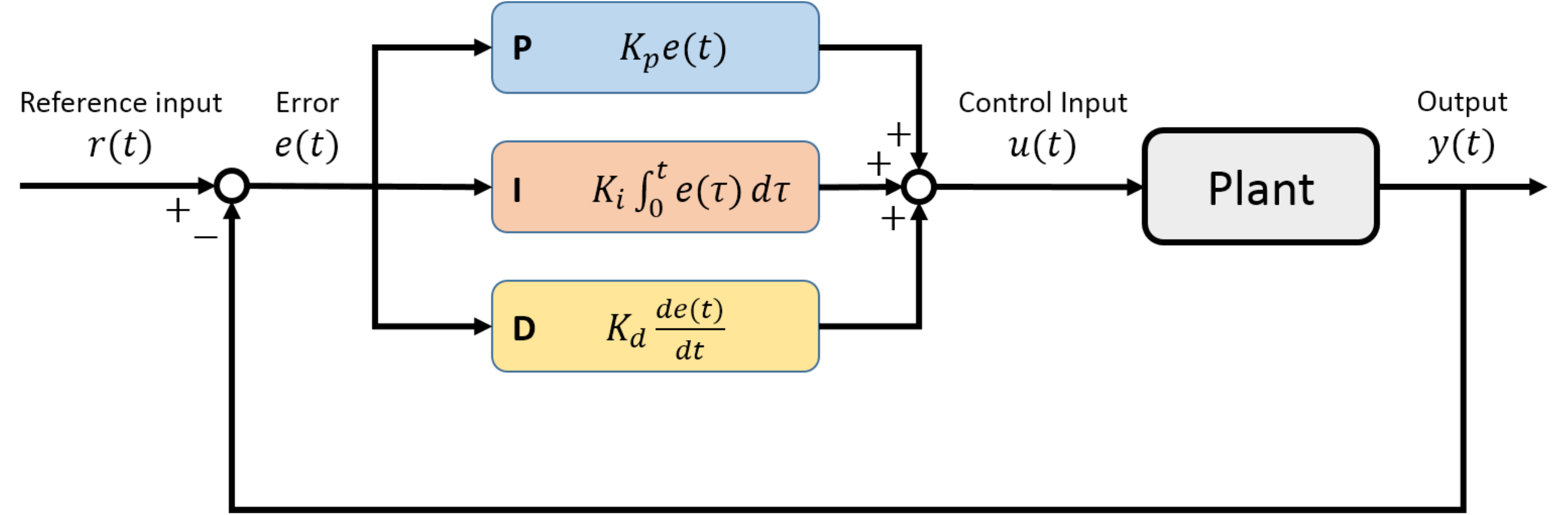}
\caption{Block diagram of PID controller.}
\label{fig_pid}
\end{figure}

\subsubsection{Proportional Integral Derivative Controller} \label{pid}
The PID controller (Fig. \ref{fig_pid}) is one of the most popular controllers due to its simplicity. PID controllers are considered a classical approach in control theory, and it is used more often than not for numerous mechanical and electrical systems. Not surprisingly, PID controllers are widely utilized having advantages over more complicated formulations. Some of the more attractive advantages include: PID controllers are very easily to implement, the parameters (gains) are easy adjusted, and PID algorithms yield a very reliable and consistent performance. However, since quadrotors are nonlinear under-actuated systems \cite{kim2016comparison}, it is not always suitable to implant the PID control directly for the quadrotor system. Nonetheless, many researchers adopted the PID controller specifically for quadrotor systems. Due to this effort, PID controllers are widely being used for many of the commercial quadrotor systems nowadays.

\begin{figure}[!t]
\centering
\includegraphics[width=0.3\textwidth]{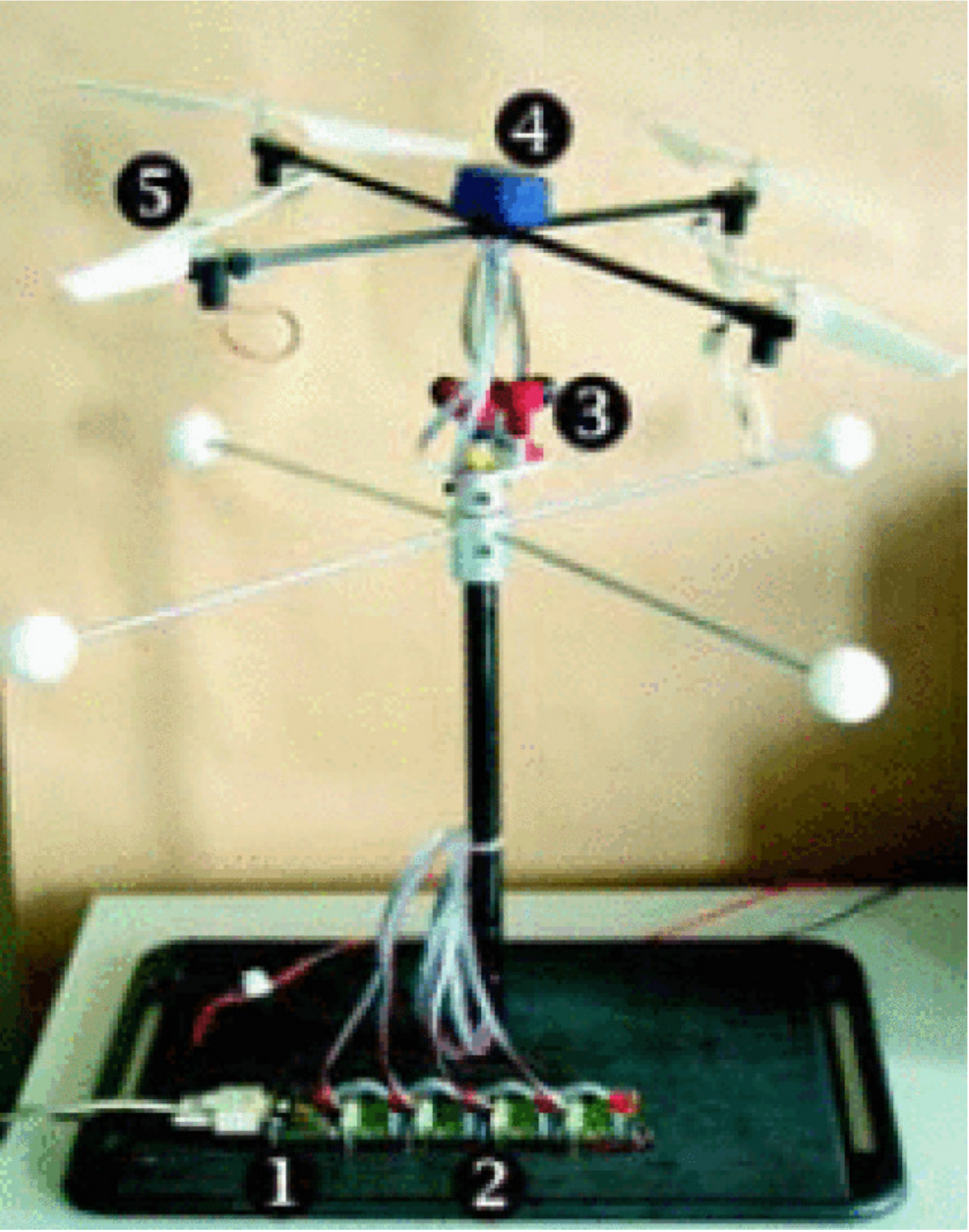}
\caption{The Ecole Polytechnique F\'{e}d\'{e}rale de Lausanne (EPFL) quadrotor test-bench for control strategies \cite{bouabdallah2004design}.}
\label{fig_bouabdallah2004design}
\end{figure}

In \cite{bouabdallah2004design, bouabdallah2004pid, bouabdallah2005towards}, S. Bouabdallah et al. designed the PID controller for the fully autonomous quadrotor. They derived the dynamic system model of the quadrotor using Euler-Lagrange formulation including the gyroscopic effects. As a next step for experiments, the PID controller was applied to a quadrotor which shown in Fig.~\ref{fig_bouabdallah2004design}, and the results were compared with a Linear Quadratic (LQ) controller. Experimental results proved that the PID controller successfully controlled the orientation angles under the perturbation and obtained the positive perspective towards AQ.

G. M. Hoffmann et al. addressed issues that arise when the flight regime deviates significantly from hover \cite{hoffmann2007quadrotor}. Specifically, they addressed three separate aerodynamic effects; namely, velocity, angle of attack, and airframe design. After some theoretical development they validated their work through test stand measurements and experimental vehicle flight tests. During their test they discovered that a PD controller was sufficient to control the vehicle in pitch maneuvers, but as speed increased blade flapping required additional measures. However, position control was successfully implemented using a PID controller that stabilized roll and pitch control inputs. Their results showed that existing models and control techniques were insufficient when used for tracking at high speed and in uncontrolled environments where wind and other issues might arise. The work showed that improved autonomous hover and trajectory tracking is possible with quadrotor of the Stanford Testbed of Autonomous Rotorcraft for Multi-Agent Control (STARMAC) project.

T. Zhang et al. developed a control architecture for autonomous hovering based on marker design, image processing, an IMU, and pose estimation and control theory \cite{zhang2009autonomous}. They used pose estimation feedback and four PID controllers with a closed-loop system to obtain autonomous hovering at altitude. Their work was also experimentally verified in real time at altitude hovering over markers with some small oscillations. This paper validated the control strategies through simulations and experiments. 

Sadeghzadeh et. al. further extended the basic PID controller to include fault tolerant control \cite{sadeghzadeh2011fault}. In their work they compared a Model Reference Adaptive Control (MRAC) and a Gain-Scheduled PID (GS-PID) model for damage. They assessed their work based on partial damage of one of propellers during flight experimentally. The GS-PID and MRAC methods were shown capable to compensate for fault/damage during hovering and flight conditions. However, the best results were obtained by combining a MIT rule that uses a Linear Quadratic Regulator (LQR) technique with the MRAC. Nonetheless, both control techniques proved successful in compensating for damage to the propeller blades.

A PID controller was also applied on the quadrotor using pose and wrench control algorithms in \cite{bellens2012hybrid}. In this work the authors used a hybrid control framework for quadrotors capable of using both pose and wrench control for task definition. Their method allowed easy switching between pose and wrench control modes. Their work was further verified experimentally by simultaneously executing forces on one direction while moving in another.

In \cite{goodarzi2013geometric}, F. Goodarzi et al. proposed a nonlinear PID controller that followed attitude and position tracking command assuming uncertainties exists on translation and rotation dynamics of the quadrotor. This controller was developed using a special Euclidean group and contained a new integral term which provided asymptotic convergence of tracking errors while there are uncertainties in the quadrotor dynamics. The simulation and preliminary experimental results demonstrated the methods utility.

H. Yang et al. proposed a PD controller for a quadrotor using a dual closed-loop control framework in \cite{yang2018active}. In the proposed framework, active disturbance rejection control and PD control strategies are applied to the inner and outer loops, respectively. From the inner loop, the perturbations of gust wind are estimated and both convergence and stabilization are given for the closed-loop system, and simple PD control strategy is proposed for control of attitude angles. Using the Lyapunov theory, the stabilization of the inner and outer closed-loop system was proved. The experimental results validate the proposed controller has the advantages in dealing with wind disturbances.

\subsubsection{Linear Quadratic Controller} \label{lqc}
In this section we examine two types of Linear Quadratic Controllers, namely: Linear Quadratic Regulator (LQR) controllers and Linear Quadratic Gaussian (LQG) Controllers for quadrotor control. In the former, the system is optimized based on a cost function and minimum cost by weighting factors supplied by the user. In the latter, the LQG controller is a combination of a Kalman type filter with a linear-quadratic regulator LQR. In the following we look at several implementations of these methods for AQ.

In \cite{bouabdallah2004pid}, Bouabdallah et. al. presents the results of two model-based control techniques, namely PID and LQ controllers for quadrotor aircraft. The PID controller showed the ability to control the orientation with small perturbations. In order to obtain the LQ controller, their system was linearized and the Riccati equation was then solved using two methods; the Pearson method and the Sage-Eisenberg method. Since finding weight matrices, which critically affects stability of the quadrotor, is problematic, the simulation and experiment results showed marginal results. However, it proved that a reliable quadrotor control was possible with an appropriate weight matrices. Some of the difficulties were attributed to the systems inaccurate model that neglected some of the couplings in the LQ controller. In addition, there were perturbations from the tethering system and other inputs causing additional difficulties.

\begin{figure}[!t]
\centering
\includegraphics[width=0.45\textwidth]{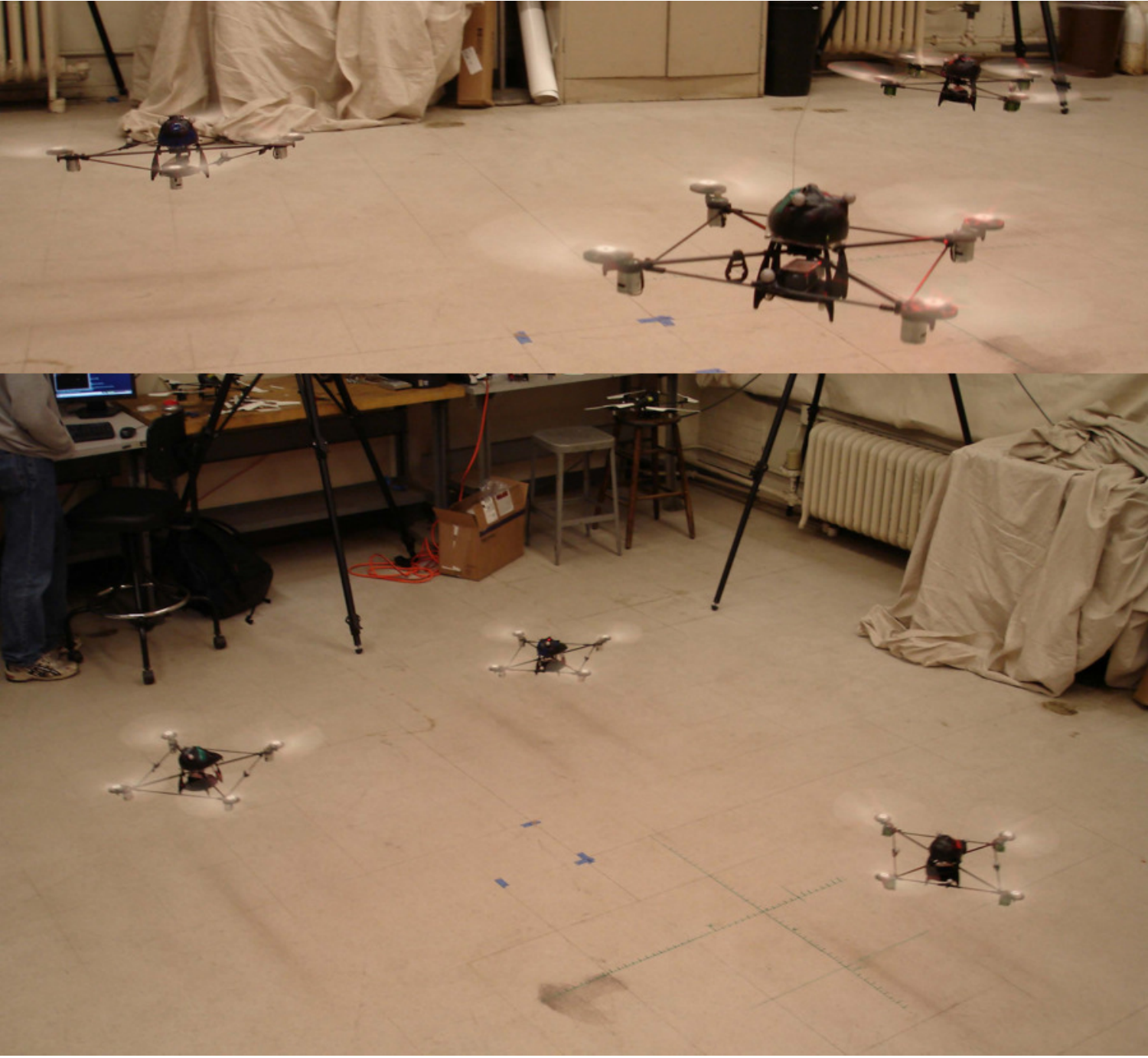}
\caption{Multiple quadrotors flight test developed at MIT \cite{valenti2006indoor}.}
\label{mit_raven}
\end{figure}

In two papers, \cite{valenti2006indoor} and \cite{how2008real}, Bethke et al. developed techniques using LQR to perform flight tests indoors for long duration missions. The work focused on issues related to single and multi-vehicle health management. These included, vehicle failures, refueling, and routine maintenance. Ultimately, the goal was to improve the overall mission performance, with vehicles that are prone to failures. Their work applied LQR technique to control quadrotors and discussed testbed infrastructure while presenting flight results using single and multi-vehicle surrogates (Fig.~\ref{mit_raven}).

B. Yu et.al. compared two control algorithms: one based on LQR techniques and the other based on a Model Predictive Control (MPC) \cite{yu2013fault}. Both methods were subjected to scenarios with and without an actuator fault. To design LQR, the static error was considered with the discrete formulation from using a digital computer. In addition, the feedback control law was considered to make quadrotor follow a reference input under the environment without static errors. Performance was evaluated based on simulations with a specific quadrotor in mind. The simulations showed that an acceptable performance can be obtained for both fault free and with actuator fault using both in a quadrotor.

\subsubsection{$H_\infty$ Controller} \label{h_infinity}
A system with external disturbances and model uncertainties, the $H_\infty$ controller can be a good choice for linear control of quadrotors. The quadrotor system is routinely affected by wind gusts and model uncertainties. Thus, some researchers have applied the $H_\infty$ controller into the quadrotor system to make the system more robust to external disturbances.

To use the $H_\infty$ controller, the control problem is typically described as a mathematical optimization problem. Basically, the $H_\infty$ controllers are used to formulate controllers that will achieve stabilization with good performance. The $H_\infty$ methods, expresses the control problem as a mathematical optimization problem. Using the $H_\infty$ technique the formulation derives a controller using the Riccati equations thereby solving an optimization problem to control the quadrotor. To do this the Linear Matrix Inequality (LMI) approach has typically been applied for solutions of the Riccati equation. On the other hand, the nonlinear $H_\infty$ controller is generally obtained by Hamilton-Jacobi equations which can replace the linear Riccati assumptions.

M. Chen et al. designed $H_\infty$ linearized controller to control the velocities, throttle and yaw for the quadrotor in \cite{chen2003combined}. The $H_\infty$ controller was combined with the model-based predictive controller to improve the constraint handling ability when the quadrotor makes aggressive maneuvers. To solve the trajectory tracking problem, the proposed $H_\infty$ controller was optimized for stabilization of the velocities, throttle and yaw control, using a loop shaping technique. The simulation showed that the combination of the two methods will work for a variety of different trajectory scenarios.
 
\cite{mokhtari2005robust} presented a generalized $H_\infty$ controller using feedback linearization for the actuator saturation. Using this approach, the controller successfully accomplished tracking reference inputs with a nonlinear quadrotor system. The performance of the proposed controller tracked error trajectories in the simulation and showed that the quadrotor system can account for disturbances and uncertainties of mass and inertia with limited actuator saturation.
 
G. V. Raffo et al. proposed a $H_\infty$ control method for solving the path tracking problem in \cite{raffo2008backstepping, raffo2010integral}. These two papers dealt with the control structure which is based on a nonlinear $H_\infty$ controller to achieve path following in the presence of external disturbances and modeling errors. In \cite{raffo2008backstepping}, the nonlinear $H_\infty$ controller was combined with the backstepping control strategy for the rotational movements and reference trajectory tracking. The robustness of $H_\infty$ was shown in the simulation. In \cite{raffo2010integral}, a mixed structure of a Model Predictive Controller (MPC) and the nonlinear $H_\infty$ controller was proposed, and MPC was utilized as a part for tracking the reference trajectory comparing to \cite{raffo2008backstepping}. In this simulation improved tracking performance and robustness was achieved with the proposed controller when compared with the results of backstepping controller.

\subsection{Nonlinear Controller} \label{nonlinear}
In this subchapter, some nonlinear controllers for the quadrotor are reviewed. Since the quadrotor system has four inputs and six degrees of freedom, it can be considered as a nonlinear under-actuated system. Therefore, to get better performance, a nonlinear controller is warranted. Since the control of quadrotors has been looked at for a number of years now, there are a large body of papers developing nonlinear control theory for the AQ. These approaches include, feedback linearization, backstepping control techniques, and sliding mode control to mention only a few.

\subsubsection{Feedback Linearization} \label{feedback}
One of the more common approaches in nonlinear control is feedback linearization. Using this method the nonlinear system is transformed into an equivalent linear system. Then using the linear systems, similarity transformation were used to produce a nonsingular matrix. This is a form of diffeomorphism and can be used to transform the state variables of the nonlinear system into a linear system. Then a standard linear control theory can be applied to the system and subsequently the solution from the linearized system is converted back into the nonlinear system.

\begin{figure}[t]
\centering
\includegraphics[width=0.48\textwidth]{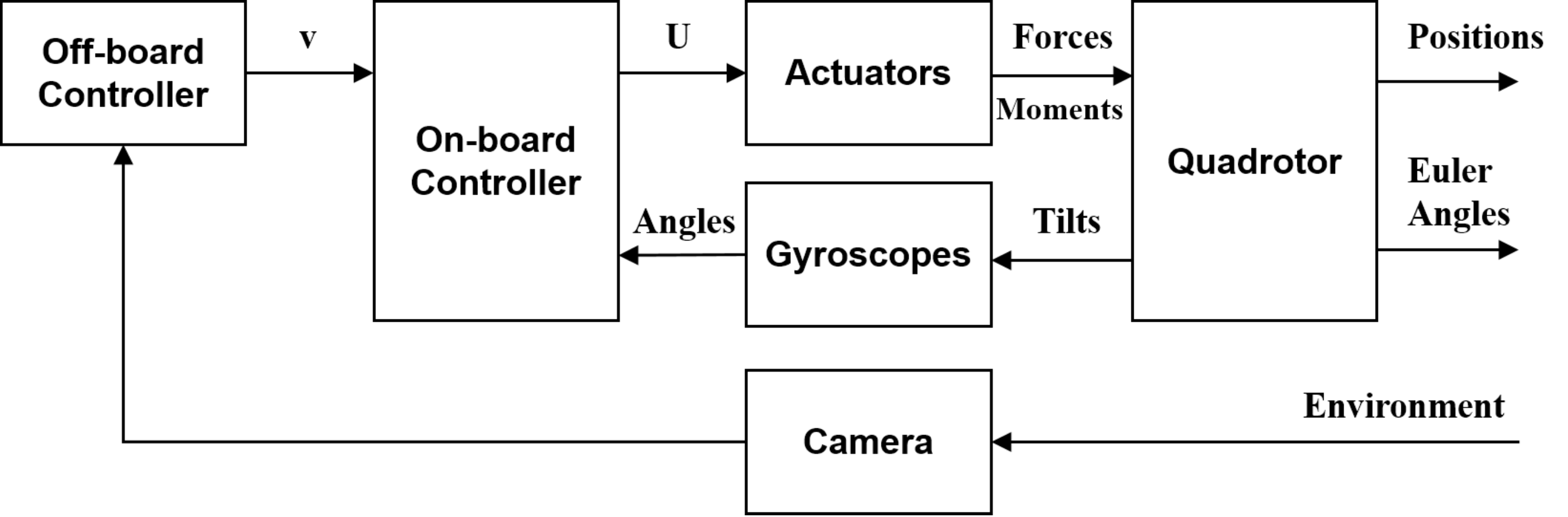}
\caption{Architecture of control system developed by E. Altug et al. \cite{altug2002control}.}
\label{fig_erdic2002control_01}
\end{figure}

E. Altug et al. presented quadrotor control methods using visual feedback as the primary sensor \cite{altug2002control}. Their work included feedback linearization and backstepping controllers and used Matlab Simulink to prove out their efforts. The architecture of control system used in their work is described in Fig.~\ref{fig_erdic2002control_01}. In the simulations, they discovered that the backstepping controller performed better than the feedback linearization. They additionally discovered that their external vision system was insufficient for fully autonomous control due to limitation of one camera which provides only good control performance of yaw and height. To achieve full autonomous control, they proposed future work where onboard and one more camera on the ground will be used together to minimize the errors of estimated tilt angles.

In \cite{lee2009feedback}, D. Lee et al. presented two types of controllers for AQ using nonlinear techniques. These included a feedback linearization controller that involved high-order derivative terms. They showed that this method was sensitive to sensor noise and modeling uncertainty. To avoid complicated derivative terms from the repeated differentiation, related with the sensitivity to noise, this paper assumed that there is the small-angle variation while ignoring the outer disturbance. The gains of the feedback linearization controller were calculated from the LQR method to demonstrate the simulation flight, and the result without any disturbances. However, since higher-order derivative terms of states and the exact dynamic equations are required to obtain the inputs of the feedback linearization controller, the proposed controller was sensitive to disturbances. They showed that this method was sensitive to sensor noise and modeling uncertainty. Then D. Lee et al. also proposed the adaptive sliding mode controller considering some uncertainty and sensor noise. (See \ref{smc})

H. Voos et al. presented a control system for a quadrotor based on combining control strategies \cite{voos2009nonlinear}. These included feedback linearization that dealt with the nonlinear dynamic behavior. The developed model was simulated using Matlab and Simulink. The proposed control method was subsequently tested in a quadrotor system. The simulation results showed satisfying attitude control performance using the proposed feedback linearization technique.

\subsubsection{Backstepping} \label{backstepping}
Backstepping is a technique for a special class of nonlinear dynamics systems. The technique is constructed from subsystems that can be stabilized using other methods. The process starts with a known-stable system and "back out" new controllers that progressively stabilize each of the subsystems. The process completes when the final control is achieved.

\begin{figure}[!t]
\centering
\includegraphics[width=0.45\textwidth]{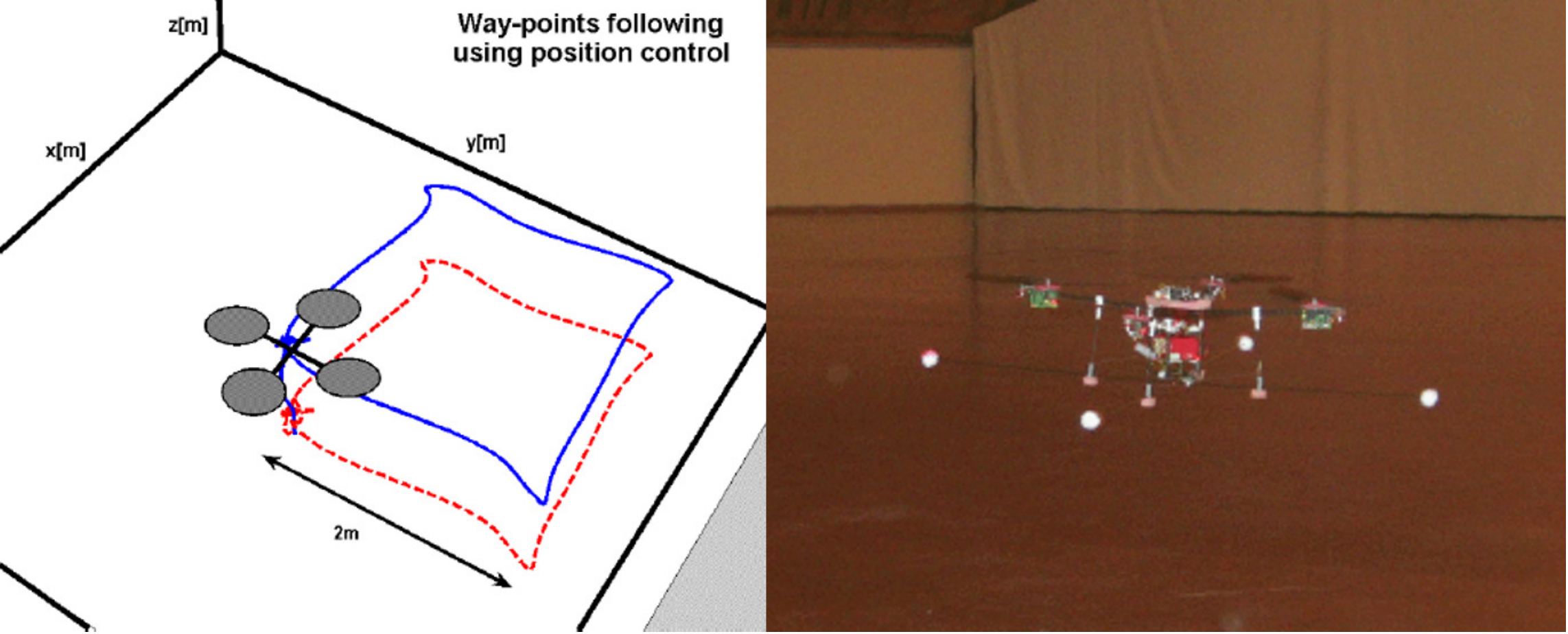}
\caption{Position control and hovering by the OS4 quadrotor \cite{bouabdallah2007full}.}
\label{fig_ethz}
\end{figure}
 
S. Bouabdallah et al. presented applications of control strategies into the OS4 quadrotor using the backstepping technique in \cite{bouabdallah2005backstepping, bouabdallah2007full}. In \cite{bouabdallah2005backstepping}, the backstepping control of the quadrotor is proposed using position tracking-errors and the Lyapunov theorem. With this approach, three inputs are extracted for control of the rotation, and one input is obtained for the altitude controller. After that, they developed the improved controller using an integral backstepping approachin \cite{bouabdallah2007full}. For the integral backstepping control design, the angle tracking-errors and its dynamics are considered using integral terms. Finally, the derived equations contained control inputs, so that control inputs were able to be obtained with the desirable dynamics for the attitude control. Also, the altitude and position controllers were obtained using the same strategy. Both papers presented results of simulations and experiments. While the proposed controller in \cite{bouabdallah2005backstepping} showed successful following results of the target position and heading angle, the proposed control strategy in \cite{bouabdallah2007full} achieved good results for not only position control, but also attitude and altitude controls by using the OS4 quadrotor shown in Fig.~\ref{fig_ethz}. In \cite{bouabdallah2007full}, the proposed controller for the quadrotor was demonstrated in an autonomous flight using a sonar-based obstacle avoidance system as well.

\begin{figure}[t]
\centering
\includegraphics[width=0.48\textwidth]{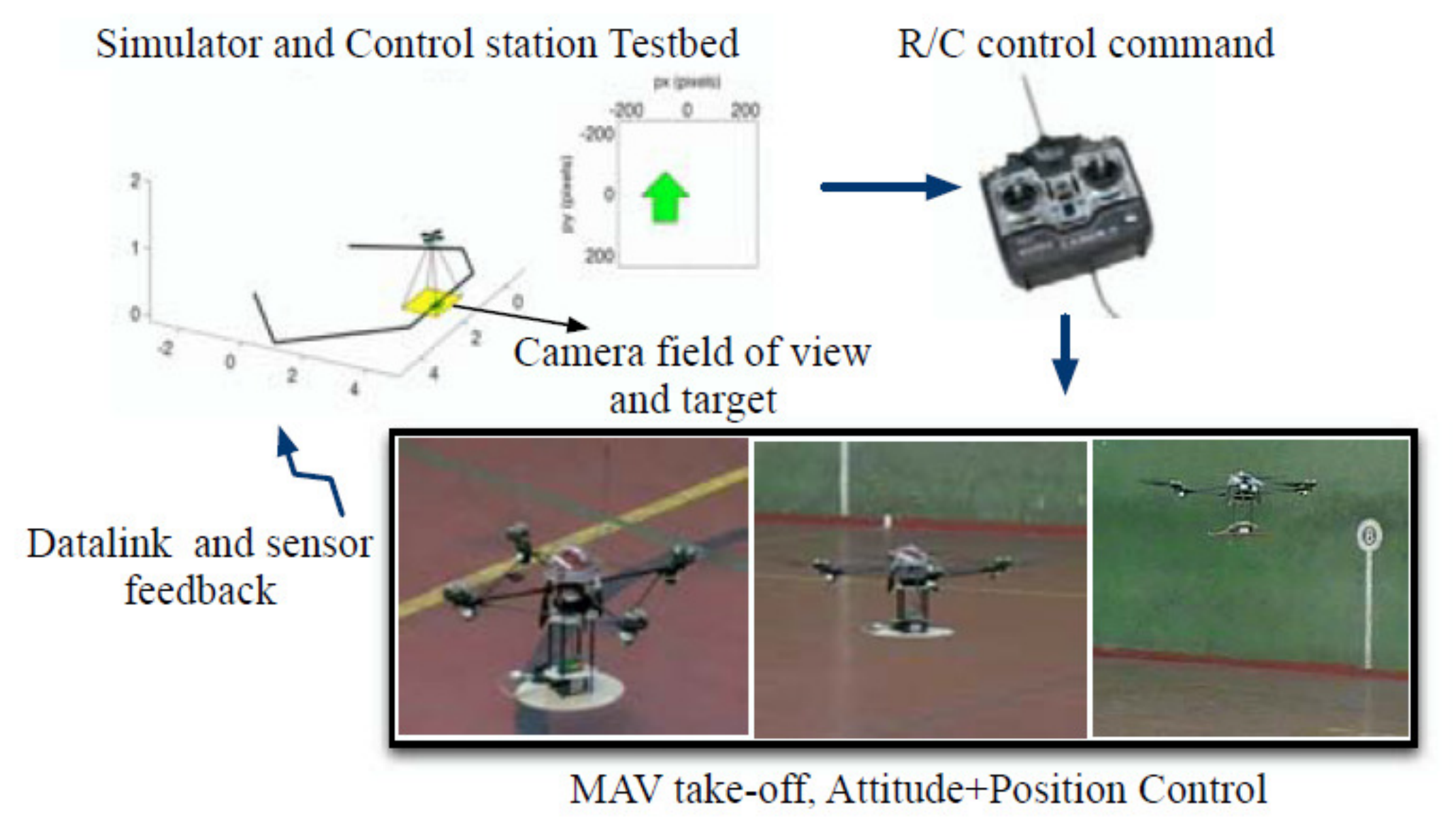}
\caption{Matlab testbed and experimental system for testing autonomous control of the quadrotor based on tracking a ground target \cite{colorado2010mini}.}
\label{fig_colorado2010mini_01}
\end{figure}

In \cite{colorado2010mini}, backstepping and the Frenet-Serret theory are used for control of quadrotor's attitude with consideration of blade flapping. The hybrid backstepping control technique proved useful for altitude stability of the quadrotor. Simulation and experiments were conducted to validate the performance of the proposed controller. Figure~\ref{fig_colorado2010mini_01} shows the Matlab testbed and experimental system for testing autonomous control of the quadrotor. The results were compared with the result of a standard PID controller, and they showed that the backstepping controller with the Frenet-Serret theory reduced the position tracking error by compensating angular acceleration immediately.

V. Lippiello et al. published a paper about the emergency landing of the quadrotor with the propeller failure \cite{lippiello2014emergency}. To do this the method turns off the opposing rotor to the broken one, and the quadrotor was considered as a birotor. For this problem, the backstepping approach was proposed. Consideration of birotor control made the quadrotor follow the planned emergency path well. However, its performance showed that yaw and roll angles were uncontrollable, but bounded using this control methodology. Simulation results of the proposed controller yielded some path following performances in various cases. One of these simulations conducted using an obstacle and its results were encouraging. Future experiments were also planned.

\subsubsection{Sliding Mode Controller} \label{smc}

\begin{figure}[!t]
\centering
\includegraphics[width=0.40\textwidth]{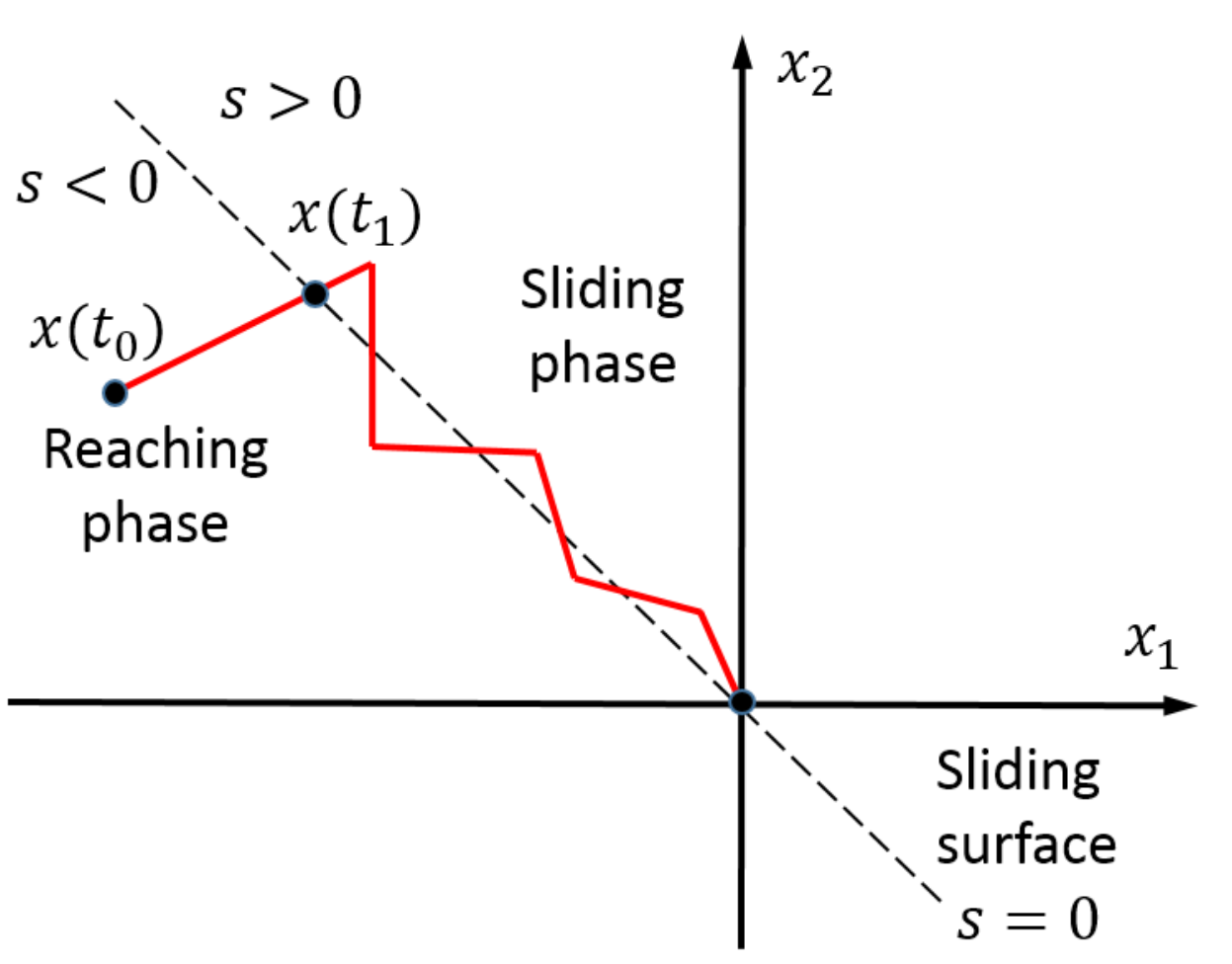}
\caption{Chattering behavior due to delay in control switching.}
\label{fig_chattering}
\end{figure}

Sliding Mode Controller (SMC) is a nonlinear control method that modifies the system using a discontinuous control signal thereby forcing the system to move within the system's normal behavior \cite{utkin1977variable}. The control law is not continuous in time and it switches from one state to another based on position in state space. While there are model uncertainties and external disturbances, this control technique guides the system to the sliding surface. The sliding surface is located between the control structures, so that the control law has to switch from one structure to another one. Hence, SMC technique can be classified as one of the variable structure control method. This characteristic of the control law and delay in control switching is a disadvantage when using SMC causing some chattering behavior as shown in Fig. \ref{fig_chattering}. To reduce the chattering, some ideas have been suggested. The first idea is to design the switching components control law in the continuous control one for reducing the amplitude of chattering. The second idea is to approximate the signum function by the saturation function which has a high slope. To design the sliding mode controller, a designer defines a sliding surface, and then designs the controller for the reaching phase where the system stay on the sliding surface. In the reaching phase, the controller can be proposed by Lyapunov theory which assures stable conditions on the sliding surface in finite time.

 S. Bouabdallah et al. applied the SMC into the quadrotor in \cite{bouabdallah2005backstepping}. In this paper, the SMC was designed for the rotations subsystem control. The sliding surface was defined and verified using the Lyapunov theory. At the final step, control inputs were extracted using a backstepping approach. This paper gave results of simulation and experiments with the designed controller. The proposed SMC was compared with the backstepping controller, and demonstrated that the backstepping controller was better than the SMC because of the switching nature of the controller with high frequency gain.
 
R. Xu et al. proposed the SMC for the stabilizing of the quadrotor under model error, parametric uncertainties, and other disturbances in \cite{xu2006sliding}. The goal of the proposed controller was to make the quadrotor fly to the desired position with a desired heading angle. Furthermore, the proposed controller used the continuous approximation of the sign function to avoid the chattering. The simulation results presented that the proposed controller successfully achieved their goal with acceptable outputs even when the simulation was conducted with parametric uncertainties.

In \cite{lee2009feedback}, D. Lee et al. presented the adaptive sliding mode controller for the quadrotor as an alternative for the feedback linearization controller. The appropriate sliding surface was defined and adaptation rules were considered to design the controller that made the quadrotor follow the desired trajectory in the presence of ground effect and sensor noise. Since the quadrotor is an under-actuated system, the augmentation of slack variables was used for designing the sliding mode controller. The proposed controller used the noise filter and saturation function to achieve the successful performance under the sensor noise, and also the adaptation rule was applied for countervailing the ground effect. The simulation and experimental results showed that the proposed adaptive sliding mode controller achieved successful control performance of the quadrotor. These results were compared with the results of the feedback linearization controller and it was confirmed that the adaptive sliding mode controller performed better than the feedback linearization controller under the noise and disturbance.

R. L{\'o}pez-Guti{\'e}rrez et al. presented the adaptive sliding mode controller combined with a robust attitude control in \cite{lopez2017robust}. The proposed controller introduces an adaptation rule in the control law to decrease the gain preserving minimal control input and holding the property of a finite-time convergence. All the simulation and experimental results show the effectiveness of the proposed controller with external disturbances and determined that the proposed controller reduced the chattering amplitude by minimizing the gain.

\subsection{Intelligent Controller} \label{intelligent}
This chapter handles intelligent, model predictive, fuzzy logic, and neural network controllers. One unique characteristic of an intelligent control is that it covers a very wide range of uncertainty compared with other control strategies. This reason led to development of control strategies such as mode predictive, fuzzy logic, and neural network controllers. The following subsections deal with these controllers and application to the quadrotor.

\subsubsection{Model Predictive Controller} \label{model}
By increasing the required coverage of uncertainty, the control strategies need to predict the future behavior of the system and generate the future control input for optimizing a cost function. The model predictive controller (MPC) is categorized as an advanced process control method that is used for maintaining the output at the operational conditions and set points. The MPC strategy is particularly suited for problems with constraints on input, output and states, and varying objectives and limits on variables. Although the MPC needs a precise prediction model and full-state estimation, it has the advantage that it can enforce constraints on inputs and outputs, and its systematic design is easy to maintain.

K. Alexis et al. suggested a switching model predictive controller (SMPC) for the quadrotor \cite{alexis2011model}. The proposed SMPC controlled the altitude and position of the quadrotor, so that the quadrotor performed indoor flight using IMU, sonar and optical flow sensor under the GPS-denied environment. They built their experimental system applying sensors to the quadrotor and achieved stable flight with this system. The experiment was designed for holding the position and showed that the proposed control strategy yielded satisfactory result.

\begin{figure}[!t]
\centering
\includegraphics[width=0.48\textwidth]{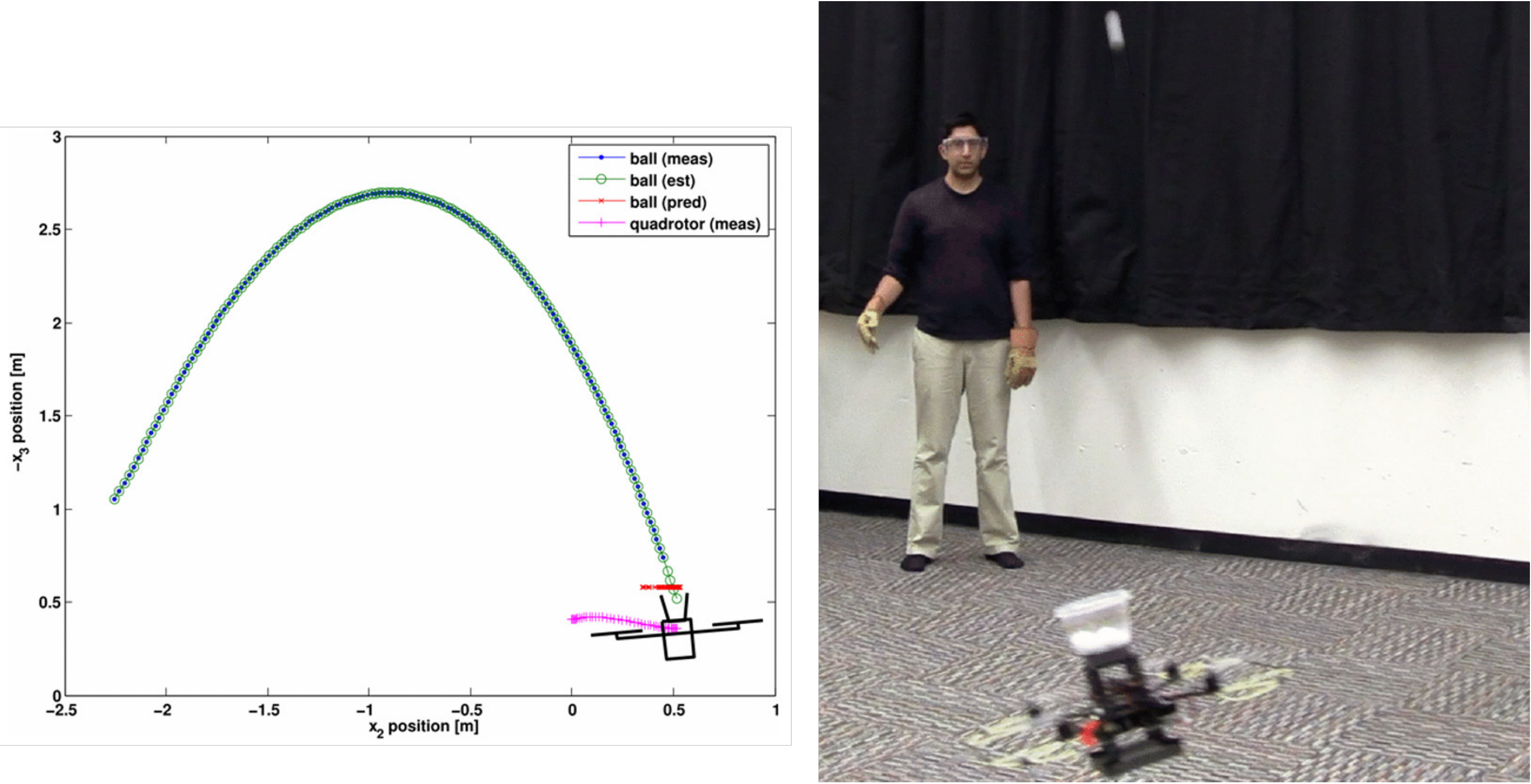}
\caption{The quadrotor is catching a ball estimating the trajectory and the final position of the ball \cite{bouffard2012learning}.}
\label{fig_bouffard2012learning}
\end{figure}

A. Aswani et al. introduced learning-based model predictive controller (LBMPC) for the quadrotor in \cite{aswani2013provably} and \cite{aswani2012extensions}. In \cite{aswani2013provably}, LBMPC was proposed for safety of the system under reasonable conditions. The proposed strategy enabled safety and performance to be decoupled maintaining two models of the quadrotor system. The uncertainty of the system was handled using statistical method. The LBMPC ensured robustness and enhanced the performance index by minimizing a cost function with selected inputs. The LBMPC was restricted in sustaining three types of robustness only. To overcome drawbacks of the LBMPC, a modified LBMPC was proposed in \cite{aswani2012extensions}, and using this strategy a continuous value function was imposed. The method assumed that the noise enters during the first time step. The proposed strategy obtained widely expanded regions of feasibility for the optimization problem. The experimental results showed that the response of the proposed LBMPC is quicker than the linear MPC. In another experiment, the quadrotor maintained flight during drastic changes in altitude. The experimental results are given in \cite{bouffard2012learning}. The experimental scenario was to catch a ball, thrown with an unknown trajectory, using the quadrotor, and verify the proposed control strategy with respect to learning of ground effect as shown in Fig. \ref{fig_bouffard2012learning}.

In \cite{yu2013fault}, B. Yu et al. proposed the MPC for the quadrotor under the actuator faults situation. When an actuator of the quadrotor failed, the proposed MPC was provided for the control of height. The simulation results of the proposed MPC showed that the performance of the proposed controller is better than LQR in the presence of a faulty actuator. In \cite{abdolhosseini2013efficient}, M. Abdolhosseini et al. developed MPC using an efficient MPC (eMPC) algorithm. Through a model reduction and MPC, they achieved a reduced number of prediction points and less computation demands. Also, the simulations of following trajectory were performed and validated in this paper, and the experimental results showed that the proposed strategy yielded satisfying performance with the quadrotor.

\subsubsection{Fuzzy Logic} \label{fuzzy}
In a fuzzy logic control theory, the term of fuzzy logic is defined as a many-valued logic in which there are more than two true values of variables, in contrast to classical or digital logic. The basic concept of the fuzzy logic was proposed in \cite{zadeh1965fuzzy} and this has been developed through numerous studies thus far. The powerful advantage of fuzzy logic is that the solution can be expressed in terms that a human can understand, unlike neural networks and genetic algorithms.

\begin{figure}[!t]
\centering
\includegraphics[width=0.45\textwidth]{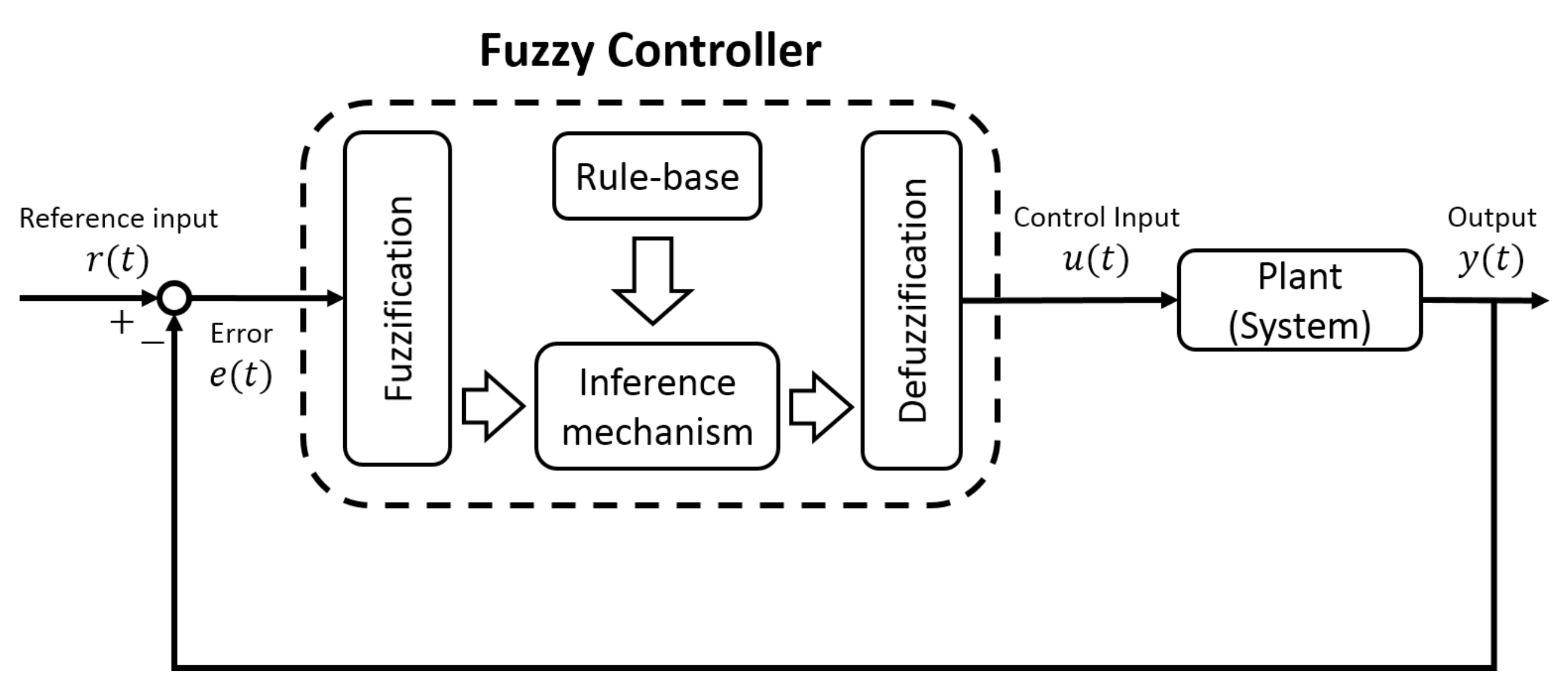}
\caption{Block diagram of simple fuzzy logic controller.}
\label{fig_FLC_block_diagram}
\end{figure}

One of applications of fuzzy logic is that the fuzzy logic controller defines a static nonlinear control law using fuzzy rules. Basically, fuzzy control rules are a series of if-then statements and provide a simple expressing control policy and domain knowledge. The if-part describes a fuzzy region in the state space and the then-part specifies a control law. To determine control input, fuzzy controller processes three steps which consist of fuzzification, rule evaluation and defuzzification. The block diagram of simple fuzzy logic controller is illustrated in Fig. \ref{fig_FLC_block_diagram}. Each component in Fig. \ref{fig_FLC_block_diagram} can be explained as follows:

1)	A Rule-Base (a set of If-Then rules) holds the knowledge, in the form of a set of rules, of how to achieve the best control result.

2) An Inference Mechanism evaluates control rules related with the current time and then decide what the best input to control the plant.

3) A Fuzzification interface simply modifies the control inputs into information that the inference mechanism can easily utilize to compare to the rules in the rule-base.

4) A Defuzzification interface converts the results of the inference mechanism into actual inputs for the plant.
Fuzzy logic has been used in a wide variety of applications in engineering, science, business, medicine, psychology, and other fields. In this subsection, we will look over the application of fuzzy logic in AQ.

In \cite{coza2006new}, C. Coza et al. proposed a new robust adaptive-fuzzy control method for a quadrotor stabilization. The proposed controller, which considered the regulation problem in the presence of oscillating wind disturbance, prevented the drift of the fuzzy membership function centers which is caused by insufficient updates. Whereas a sliding mode controller cause chattering of control signal and the neural network controller was computationally intensive when using an an on-board computer. The proposed adaptive fuzzy controller showed good stability in the simulation results.

M. H. Amoozgar et al. used an adaptive PID controller for fault-tolerant control of a quadrotor system that included actuator faults \cite{amoozgar2012fault}. Then a fuzzy inference scheme was used to tune the controller gains. Two fault scenarios were examined to include the total loss of actuator control loss of control in one actuator. The adaptive PID controller was contrasted with a conventional one in an experiment. The results obtained validated the usefulness of the method. 

In \cite{lee2014robust}, H. Lee et al. proposed trajectory tracking controller using Takagi-Sugeno (TS) fuzzy model. TS fuzzy model, was developed for a complex and higher-dimensional problem and reduced the number of rules required by the Mamdani model. The authors showed simulations for the stable tracking performance using the TS fuzzy model based controller. The proposed controller achieved good tracking performance in comparison with the conventional LQR controller.

\subsubsection{Neural Network} \label{neural}

\begin{figure}[!t]
\centering
\includegraphics[width=0.40\textwidth]{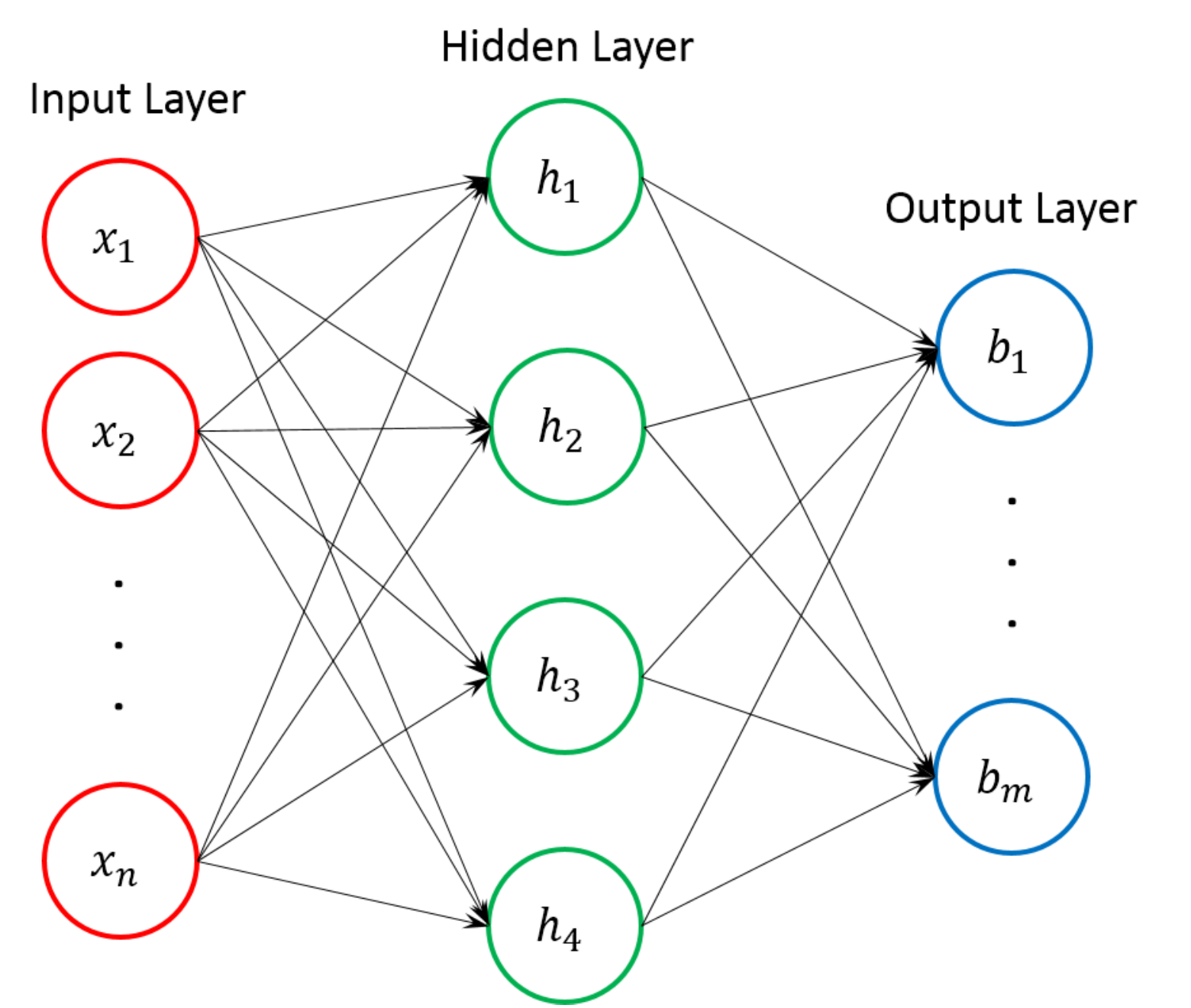}
\caption{An example of neural network system.}
\label{fig_nn}
\end{figure}

 A Neural Network (NN) is a system with the interconnection between the neurons in the different layers. In the simple example of an NN system Fig. \ref{fig_nn}, we can find three layers in which the first layer and the third layer have input and output neurons, respectively. The NN system is defined as a set of units called \textit{processing elements}. Each processing element has a weighting parameter and the behavior of the element is altered through the changing of the weights. A mathematical equation of a network structure with an input (u$\in R^n$) and output (y $\in R$) is described by (\ref{eqn_nn}).

\begin{equation}
\label{eqn_nn}
Y = f(u) = \sum\limits_{i=1}^{n} w_i g_i(u) + w_0,
\end{equation}

where $w_i$ ($i=0,1,\ldots n$) are the weights. As one of intelligent control methods, the NN control strategy has been used for a design of a nonlinear dynamic system with uncertain nonlinear terms and system errors. The objective of this control strategy is to find the weights for achieving a desired input and output. Through this process termed as \textit{training the network}, the system obtains a control law overcoming a wide range of uncertainty.

T. Dierks and S. Jagannathan have researched applications of neural network into a quadrotor \cite{dierks2008neural, dierks2009neural, dierks2010output}. In \cite{dierks2008neural} and \cite{dierks2010output}, they proposed the output feedback controller and observer based on NN for a quadrotor. The objective of this control strategy was to make the quadrotor follow a desired trajectory with uncertainties in the model and other disturbances. Using the proposed control strategy they verified that the position, orientation, velocity tracking, observer estimation, and the NN weight estimation errors were all semi-globally bounded using the Lyapunov theory. The simulation results showed the effectiveness of the proposed controller and it outperformed a conventional linear controller. They also proposed the controller based on NN for a quadrotor formation flight in \cite{dierks2009neural}. In this paper, the proposed controller allowed follower UAVs to track its leader without the knowledge of its dynamics. This novel NN control strategy was also verified by the Lyapunov theory for stability and performance.

In \cite{rinaldi2013linear}, F. Rinaldi et al. introduced the NN based control strategy for the quadrotor and applied ithe control strategy into the vertical dynamics of quadrotor. There are two steps that were performed for the system identification and control design using NN. This control strategy applied to the vertical quadrotor flight and a one-dimensional formation flight. Then the simulation was carried out and compared with the LQR strategy. The simulation results showed that the NN controller outperformed the LQR in terms of raising time, settling time and overshoot in the altitude control and formation flight control.

\section{Conclusions} \label{conclusions}
In this paper, we reviewed the basic history and control methodologies used for quadrotors. The popularity of quadrotors has sparked the imagination and ingenuity of many researchers, and has been applied to a number of different applications. This paper studied many different controllers that may be used by quadrotors in a variety of scenarios. As the number of applications grow, the need for imaginative new control strategies that build upon older ones will undoubtedly grow. This paper may serve as a guide for other researchers studying different quadrotor applications and control methodologies. Furthermore, this comprehensive survey breaks control strategies into specific categories that may help other researchers narrow down their focus and studies.


%

%
%


\ifCLASSOPTIONcaptionsoff
  \newpage
\fi



\bibliographystyle{IEEEtran}
\bibliography{IEEEabrv,mybib}
%




%

\begin{IEEEbiography}[{\includegraphics[width=1in,height=1.25in,clip,keepaspectratio]{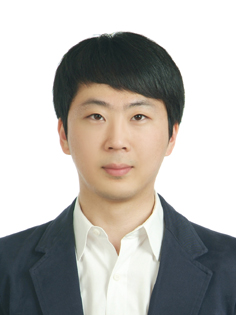}}]{Jinho Kim}
received his Ph.D. and MS degrees in Mechanical Engineering at University of Maryland, Baltimore County in 2018 and 2017, respectively. He received his bachelor’s degree in Aerospace Engineering at the Chungnam National University, South Korea in 2011. His research interests include application of nonlinear control theory, robotics, vision-based autonomous control and navigation.
\end{IEEEbiography}

\begin{IEEEbiography}[{\includegraphics[width=1in,height=1.25in,clip,keepaspectratio]{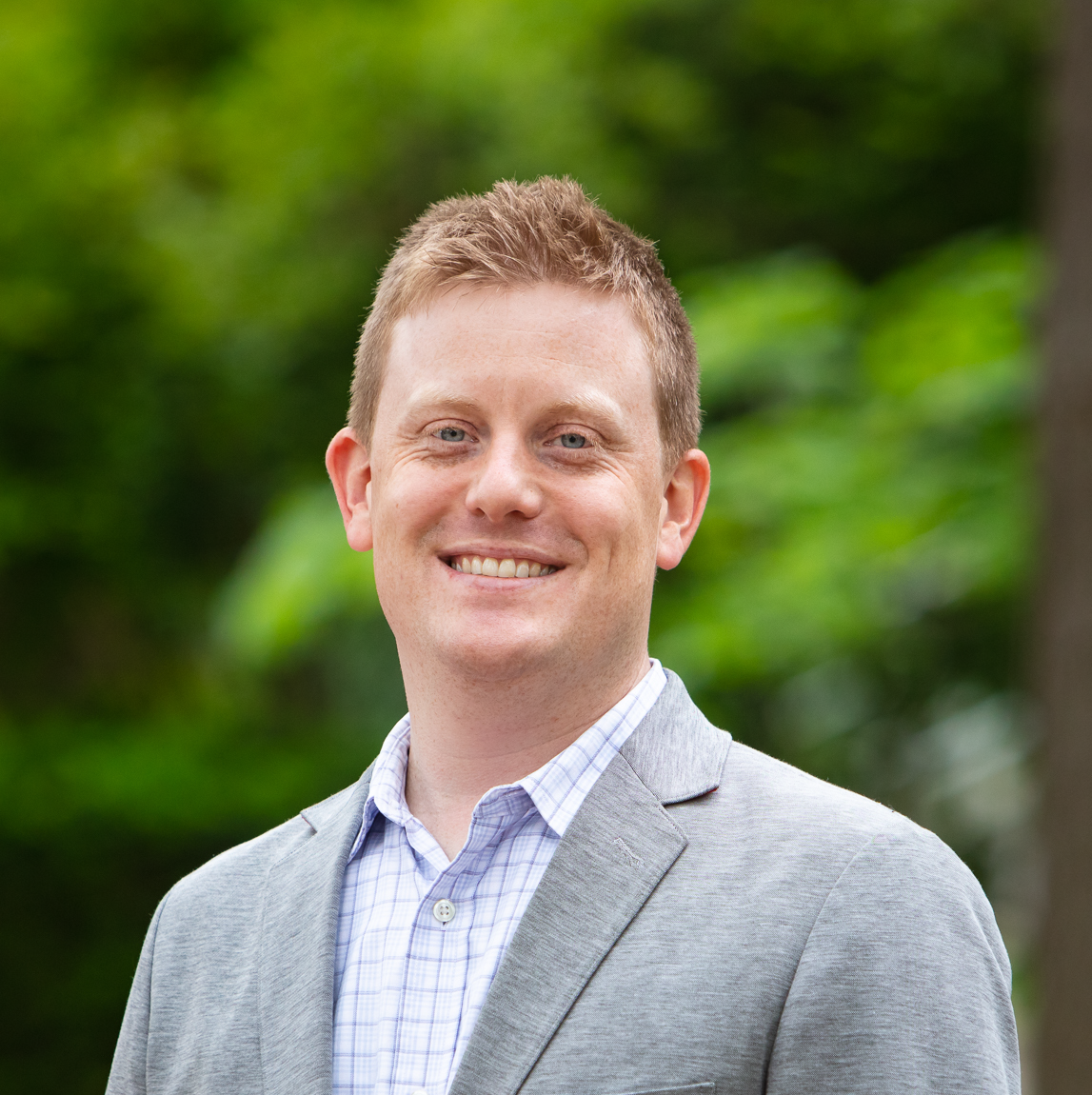}}]{S. Andrew Gadsden} received his Ph.D. in Mechanical Engineering (Thesis), McMaster University, ON, Canada, 2011 and his B.Eng.Mgt., Mechanical Engineering and Management, McMaster University, ON, Canada, 2006. He subsequently did a Postdoctoral at the Centre for Mechatronics and Hybrid Technology, McMaster University, ON, Canada, 2011-2014. He is currently an Assistant Professor at University of Guelph where he continues to mentor students and conduct his research.
\end{IEEEbiography}

\begin{IEEEbiography}[{\includegraphics[width=1in,height=1.25in,clip,keepaspectratio]{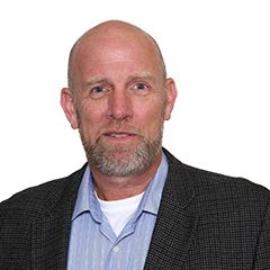}}]{Stephen A. Wilkerson} received his Ph.D., Mechanical Engineering, The Johns Hopkins University, 1990. Thesis Title: ”A Boundary Integral Approach to Three-Dimensional Underwater Explosion Bubble Dynamics.”, he also has a BS and MS degrees from John Hopkins and George Washington University in Engineering. He continues his research and mentoring of students at York College of Pennsylvania.
\end{IEEEbiography}

\end{document}